\documentclass[floatfix,aps,prd,showpacs,amsmath,nofootinbib,preprintnumbers]
{revtex4}

\usepackage{graphicx}
\usepackage{colordvi}

\begin{document}

\preprint{FERMILAB-PUB-04-145-A}

\title{Effect of inhomogeneities on the expansion rate of the Universe}

\author{Edward W. Kolb}\email{rocky@fnal.gov}
\affiliation{Particle Astrophysics Center, Fermi
       	National Accelerator Laboratory, Batavia, Illinois \ 60510-0500, USA \\
       	and Department of Astronomy and Astrophysics, Enrico Fermi Institute,
       	University of Chicago, Chicago, Illinois \ 60637-1433 USA}

\author{Sabino Matarrese}\email{sabino.matarrese@pd.infn.it}
\affiliation{Dipartimento di Fisica ``G.\ Galilei,'' Universit\`{a} di Padova, 
        and INFN, Sezione di Padova, via Marzolo 8, Padova I-35131, Italy}

\author{Alessio Notari}\email{alessio.notari@sns.it}
\affiliation{Scuola Normale Superiore, Piazza dei Cavalieri 7, Pisa I-56126 
        and INFN, Sezione di Pisa, Italy}

\author{Antonio Riotto}\email{antonio.riotto@pd.infn.it}
\affiliation{INFN, Sezione di Padova, via Marzolo 8, I-35131, Italy}

\date{\today}

\begin{abstract}
While the expansion rate of a homogeneous isotropic Universe is simply
proportional to the square-root of the energy density, the expansion
rate of an inhomogeneous Universe also depends on the nature of the
density inhomogeneities.  In this paper we calculate to second order
in perturbation variables the expansion rate of an inhomogeneous
Universe and demonstrate corrections to the evolution of the expansion
rate.  While we find that the mean correction is small, the variance
of the correction on the scale of the Hubble radius is sensitive to the 
physical significance of the unknown spectrum of density perturbations
beyond the Hubble radius.
\end{abstract}

\pacs{98.80.Cq}

\maketitle

\section{Introduction}\label{Introduction}

There is no more fundamental physical quantity in cosmology than the expansion
rate of the Universe.  In recent years the present value of the expansion rate,
Hubble's constant, has been measured with increasing accuracy
\cite{wendy}. With the exploration of the Universe at redshifts of order unity,
we now have information about the time evolution of the expansion rate
\cite{scphiz}.  A most surprising result is that the time evolution of the
expansion rate does not seem to be described by a matter-dominated
Friedmann-Lema\^{\i}tre-Robertson-Walker (FLRW) cosmological model.  The usual
explanation for the discrepancy is that there is a new component of the energy
density of the Universe, known as dark energy, that determines the recent
evolution of the expansion rate.  Of course all indications for dark energy are
{\em indirect}; they all involve some form of the time evolution of the
expansion rate.

Since the expansion rate of the Universe is of such fundamental importance, we
must understand any possible effects that would result in an expansion rate
different from the FLRW prediction. In this paper we study the change in the
expansion rate due to perturbations of a homogeneous, isotropic, FLRW model.
In particular, we perform a second-order calculation of the effect of
inhomogeneities on the expansion rate.  We only consider modifications to the
expansion rate of a matter-dominated universe, although our results can be
extended to a universe containing a mixture of matter and a cosmological
constant.  We find that the mean corrections are a few parts in $10^5$.

The expansion rate of a perturbed FLRW cosmology has been discussed in many
works (although we believe ours is the first complete second-order
calculation).  Hui and Seljak \cite{lam} estimated the order of magnitude of
the effect by considering a representative second-order term.  They arrived at
the correct mean order of magnitude of the result.  Some recent works
\cite{Bene,Rasanen} suggested that small-scale contributions could give a large
correction, producing an apparent accelerated expansion of the universe.  In
particular R\"as\"anen \cite{Rasanen} 
suggested that due to ultraviolet sensitivity, one of the 
second-order terms could give large corrections depending on boundary 
conditions. As we will discuss in the conclusions, we find that if one employs 
the correct averaging procedure the result (at least for the terms we could 
compute exactly using the second-order formalism) is well behaved in the 
ultraviolet and the term identified by R\"as\"anen does not result in a large 
correction.

In addition to calculating the mean value of the expansion rate, we calculate
the variance about the mean value.  In calculating the variance we uncover an
interesting infrared effect.  At second order the expansion rate has a term
proportional to $\varphi\nabla^2\varphi$, where $\varphi(\bf{x})$ is the
peculiar gravitational potential, related to the density perturbations through
the cosmological Poisson equation.  If inflation is the origin of
perturbations, then $\varphi$ should be a Gaussian random variable with zero
mean.  However, if the perturbation spectrum on super-Hubble-radius scales is
no bluer than a Harrison--Zel'dovich spectrum, then the variance of $\varphi$
formally has an infrared singularity.\footnote{Presumably this infrared
singularity is cut off because inflation did not last for an infinite period of
time.}  Now if the value of $\varphi$ in our local Hubble volume is
sufficiently large to modify the expansion rate, our perturbative expansion
fails.  Nevertheless, our results suggest that if the super-Hubble modes of
$\varphi$ have physical significance, then a nonperturbative extension of our
calculation could yield a most important modification to the Friedmann
equation.  One might even speculate that a complete treatment of the effect 
could explain the observed time dependence of the expansion rate on its own 
and obviate the need for the dark-energy assumption.  Since that conjecture is
beyond the perturbative calculation of this paper, we will postpone discussion
of this point to a subsequent communication \cite{kmnrtwo}.

In the next section we discuss the general perturbative expansion.  In Section
\ref{thetasynchronous} we calculate the corrections to the expansion rate in
terms of metric fluctuations, and then express the results in terms of the
density perturbation spectrum.  In Section \ref{numbercrunch} we present the
numerical results.  We then conclude, followed by two technical appendices. 

\section{The general formalism}\label{Friedmann}

In this section we describe how to treat the average properties of a
perturbed Universe up to second order in the metric variables,
including the effect of the inhomogeneous gravitational field on the
homogeneous background field.  We then discuss the proper definition
of the average expansion rate and express the effect of
inhomogeneities to second order as averages of density perturbations.

By ``average,'' we mean the average over a spatial hypersurface at a
given time. Clearly an average defined in this way depends on the
chosen coordinate system, {\it i.e.,} the gauge.  

We will consider a Universe filled only by irrotational dust and choose the
coordinates of an observer at rest with the dust ({\it i.e.,} comoving
coordinates), and with the same time coordinate for every point of the
spatial hypersurface ({\it i.e.,} synchronous coordinates). This
system of coordinates can be chosen if the Universe is filled by a
single pressureless component.  Since the pressure vanishes, the only
force acting on the particles is gravity, and the comoving world lines
coincide with geodesics. (See Ref.\ \cite{LL} for a full discussion of
this point.)

We will call $\tau$ the (conformal) time in this gauge, and $x^i$ the
spatial coordinates, so that the metric has the form
\begin{equation}
ds^2=a^2(\tau) \left[ -d \tau^2 + \gamma_{i j}(\tau,x^i) dx^i dx^j  \right]
\label{metric}
\end{equation}
We will perform the averages on constant-$\tau$ spatial hypersurfaces.

\subsection{First order}\label{firstorder}

The goal of this paper is a second-order calculation, but before
embarking, let us recall the familiar linearized first-order result.
The energy-momentum tensor and metric are expanded to first-order as
\begin{subequations}
\label{subonee}
\begin{eqnarray}
T_{\mu \nu} & = & T_{\mu \nu}^{(0)} + T_{\mu \nu}^{(1)}   
\label{defTfirstorder} \\
g_{\mu \nu} & = & g_{\mu \nu}^{(0)} + g_{\mu \nu}^{(1)} ,  
\label{defgfirstorder}
\end{eqnarray}
\end{subequations}
where the superscript $(r)$ denotes the $r$-th order perturbation.  
By definition $T_{\mu\nu}^{(0)}$ and $g_{\mu \nu}^{(0)}$ are homogeneous 
and isotropic.

In the synchronous gauge, to first order the metric may be written in
terms of a set of perturbation variables consisting of two scalars
($\psi^{(1)},\ \chi^{(1)}$), a vector ($\chi^{(1)}_i$), and a tensor
($\chi^{(1)}_{ij}$).  The vector $\chi_i^{(1)}$ is transverse
($\partial ^i\chi^{(1)}_i=0$) and the tensor $\chi^{(1)}_{ij}$ is
symmetric, transverse, and traceless
($\chi^{(1)}_{ij}=\chi^{(1)}_{ji}$, $\partial^i\chi_{ij}^{(1)}=0$,
$\chi^{(1)i}_{\ \ \ \ i}=0$).  In terms of these variables the metric
is
\begin{equation}
\gamma_{ij} = \left(1-2\psi^{(1)}\right)\delta_{ij}
+ D_{ij}\chi^{(1)} + \partial_i\chi^{(1)}_j + \partial_j\chi^{(1)}_i 
+ \chi_{ij}^{(1)} , 
\end{equation}
where $D_{ij}=\partial_i\partial_j-\frac{1}{3}\nabla^2 \delta_{i j}$.    

The metric perturbations $\psi^{(1)}$ and $\chi^{(1)}$ may be
expressed in terms of the peculiar gravitational potential
$\varphi({\bf x})$, which is related to $\delta^{(1)}$, the 
first-order density perturbation, by the Poisson equation,
\begin{equation}
\nabla^2\varphi = \frac{\kappa^2}{2} a^2\rho^{(0)}\delta^{(1)} .
\label{nabla2varphi}
\end{equation}
Ignoring metric perturbations that decay in time, for a 
matter-dominated Universe $\psi^{(1)}$ and $\chi^{(1)}$
are given by (see Ref.\ \cite{MMB})\footnote{The perturbation
variable $\phi$ in Ref.\ \cite{MMB} corresponds to our variable
$\psi$, and vice versa.}
\begin{subequations}
\label{fomp}
\begin{eqnarray}
\psi^{(1)}({\bf x},\tau) & = & \frac{5 }{3}\varphi({\bf x}) 
+ \frac{\tau^2}{18} \nabla^2 \varphi({\bf x})  \label{psi1}  \\
\chi^{(1)}({\bf x},\tau) & = & -\frac{1}{3}  \tau^2 \varphi({\bf x}) .
\label{chi1}
\end{eqnarray}
\end{subequations}
To first order we will neglect vector modes, as they do not arise in
conventional perturbation-generation mechanisms such as inflation.  We
will also assume the tensor mode amplitude is small.

In calculating spatial averages we will also require $\sqrt{\gamma}$,
where $\gamma$ is the determinant of the spatial metric.  To first
order \cite{MMB},
\begin{equation}
\sqrt{\gamma} = 1 + \frac{1}{2}\gamma^{(1)}({\bf x},\tau) = 
1 - 3\psi^{(1)}({\bf x},\tau)  = 1-5\varphi({\bf x}) - 
\frac{\tau^2}{6}\nabla^2\varphi({\bf x}) .
\label{fosd}
\end{equation}

Whether calculating to first order or second order, when calculating
the spatial average $\langle \cdots \rangle$ of a quantity ${\cal
O}(\tau,x^i)$ one must fix a system of coordinates in which to
express ${\cal O}(\tau,x^i)$, and then integrate with the proper
integration measure over $d^3\!x$ at fixed $\tau$.  For all orders we
adopt the definition\footnote{Note that this definition is appropriate for
scalar quantities. It will also be used to average $G_{00}$ which is a scalar
under the residual gauge freedom.}
\begin{equation}
\langle {\cal O}\rangle(\tau)
\equiv \frac{\int d^3\!x \, \sqrt{\gamma(\tau,x^i)} \,
{\cal O}(\tau,x^i)}{\int d^3\!x \, \sqrt{\gamma(\tau,x^i)}} ,
\label{average}
\end{equation}
where the domain of integration is some large volume.  If ${\cal
O}^{(0)}$ is a homogeneous quantity ({\em i.e.,} it does not depend on
$x^i$, but only on $\tau$), then simply $\langle {\cal O}^{(0)}
\rangle ={\cal O}^{(0)}$.  In the first-order calculation, if ${\cal
O}^{(1)}$ is already a first order quantity we may set $\gamma=1$, and
$\langle{\cal O}^{(1)} \rangle$ becomes
\begin{equation}
\langle{\cal O}^{(1)}\rangle = 
\frac{\int d^3\!x \, {\cal O}^{(1)}(\tau,x^i)}{\int d^3\!x} .
\label{ffo}
\end{equation}

To first order, the Einstein equations are
\begin{equation}
G_{\mu \nu}^{(0)} + G_{\mu \nu}^{(1)} =
\kappa^2 \left( T_{\mu \nu}^{(0)} + T_{\mu \nu}^{(1)} \right) ,
\label{foei}
\end{equation}
which yield upon averaging the $00$ component 
\begin{equation}
G_{00}^{(0)}=\kappa^2\left\langle T_{00}\right\rangle
- \left\langle G_{00}^{(1)} \right\rangle . \label{backEinstein1}
\end{equation} 
Here we see that $\kappa^{-2} \left\langle G_{00}^{(1)}\right\rangle$
may be interpreted as an extra component to the stress-energy 
tensor.\footnote{This effect is sometimes incorrectly described as a
``backreaction.'' Technically, it is not a backreaction; 
while the inhomogeneities modify the expansion rate of the Universe, 
they are not produced by the expansion of the universe.}

For the unperturbed FLRW cosmology, the expansion rate $a'/a^2$ is found from 
the $0-0$ component of the Einstein
equations.  For the first-order perturbed FLRW model 
\begin{equation}
3\left( \frac{a'}{a^2} \right)^2 = 
\kappa^2 \langle\rho\rangle - a^{-2}\left\langle G_{00}^{(1)} 
\right\rangle ,  \label{backEinstein001}
\end{equation}
where $\rho =  \rho^{(0)}+\rho^{(1)} =a^{-2}T_{00}$.  
This is the basic point.  In a perturbed FLRW cosmology,
$\dot{a}/a=a'/a^2$ is {\em not} 
$\sqrt{\kappa^2 \langle\rho^{(0)}\rangle/3}$.\footnote{Here 
the overdot stands for $d/dt$,
where $t$ is related to $\tau$ by $a d\tau=dt$. }  However,
one must be careful to find the true variables that describe the
evolution of the averaged background. It is not clear that the
quantity $a'/a^2$ describes the physical Hubble flow.  The correct quantity to
describe the Hubble flow may be found from the evolution of $\langle \rho
\rangle$.  

We know that for a homogeneous isotropic Universe there are
two independent equations that govern the dynamics of the
expansion. So we have to find the two independent equations that
describe the evolution of averaged physical quantities such as like
$\langle \rho \rangle$.
For the unperturbed model we may augment the $0-0$ equation with the
continuity equation, $D_\mu T^{\mu 0}=0$. This gives for a perfect
pressureless fluid (dust)
\begin{equation}
\frac{1}{a}\rho'=\dot{\rho}=-\theta\rho \, , \label{continuity}
\end{equation}
where $\theta\equiv D_{\mu} u^{\mu}$ and $u^\mu$ is the fluid four
velocity.  Eq.\ (\ref{continuity}) is true to all orders.  For an unperturbed 
FLRW Universe with scale factor $a$, this immediately gives 
$\rho^\prime/\rho=-3a^\prime/a$, which results in $\rho \propto a^{-3}$.

For the perturbed model, we want to expand Eq.\ (\ref{continuity}) to
first order and then average it. This will give us the effective
scale factor for the dilution of the matter density.
In addition to the expansion of $\rho$, we expand $\theta$ as
\begin{equation}
\theta = \theta^{(0)}+\theta^{(1)}    ,      
\end{equation}
 and obtain
\begin{equation}
\frac{1}{a}\left\langle\rho^{(0)\prime}+\rho^{(1)\prime}\right\rangle 
= - \left\langle \left(\theta^{(0)}+\theta^{(1)}\right)
\left( \rho^{(0)} +\rho^{(1)}  \right) \right\rangle      .        
\label{coneq1}
\end{equation}
Using the fact that $\langle\rho^{(1)\prime}\rangle =
\langle\rho^{(1)}\rangle^\prime$ we are left with
\begin{equation}
 \frac{1}{a} \frac{\langle\rho\rangle'}{\langle\rho\rangle} 
= - \langle \theta \rangle  .
\end{equation}
This tells us that in a first-order perturbed Universe the average
matter density is diluted with expansion according to
$\langle\rho\rangle^{-1} d\langle\rho\rangle/dt = -\langle \theta
\rangle$, which in general is {\it not} equivalent to
$\langle\rho\rangle^{-1} d\langle\rho\rangle/dt = -3\dot{a}/a$.  

An alternative way to express the result is to define a new scale
factor, $a_V$, such that $\langle\rho\rangle\propto a_V^{-3}$.  It is clear
that $a_V$ must be defined by
\begin{equation}
\frac{\dot{a}_V}{a_V} \equiv \frac{1}{3}\langle\theta\rangle .
\end{equation}
And so the {\it true} scale factor is $a_V$ and not $a$, at least for
the dilution of matter (see, {\em e.g.,} \cite{Buchert}).

The physical quantity of interest is $\langle\theta\rangle$. It is given by
\begin{equation}
\langle\theta\rangle = \theta^{(0)} + \langle\theta^{(1)}\rangle
= 3\frac{a'}{a^2} + \langle\theta^{(1)}\rangle .
\end{equation}
We will define a $\delta\theta$, and express $\theta$ as
\begin{equation}
\langle\theta\rangle  =  3\sqrt{\frac{\kappa^2\langle\rho\rangle}{3}} + 
   \langle\delta\theta\rangle 
=  3H\left(1+\frac{\langle\delta\theta\rangle}{3H}\right) .
\label{combo11}
\end{equation}
Two important notational points: We have defined 
$H\equiv\sqrt{\kappa^2\langle\rho\rangle/3}$ ({\it not} as $a'/a^2$);
and we have defined $\langle\delta\theta\rangle$ as the 
difference between $\langle\theta\rangle$ and $3H$ ({\it not} 
as the difference between $\langle\theta\rangle$ and $\theta^{(0)}$). 

To calculate $\langle\delta\theta\rangle/3H$, we can express 
Eq.\ (\ref{backEinstein001}) in the form
\begin{equation}
\frac{a'}{a^2} = \left( \frac{\kappa^2 \langle\rho\rangle}{3} 
- \frac{\left\langle G_{00}^{(1)} \right\rangle}{3a^2}
\right)^{1/2} \simeq H \left( 1 -
\frac{\left\langle G_{00}^{(1)} \right\rangle}{6a^2H^2}\right) ,
\label{combo21}
\end{equation}
where of course the second equality holds if the corrections are small.   
Combining Eqs.\ (\ref{combo11}) and (\ref{combo21}), we find
\begin{equation}
\frac{\langle\delta\theta\rangle}{3H} = \frac{\langle\theta^{(1)}\rangle}{3H} 
- \frac{\left\langle G_{00}^{(1)} \right\rangle}{6a^2H^2} .
\label{combo31}
\end{equation}

It will turn out that in the first-order calculation 
$g_{\mu\nu}^{(1)}$ will appear only as a spatial
gradient in the final expression for $\langle\delta\theta\rangle$.
This means that the physical result is insensitive to the choice of
the normalization of $\langle g_{\mu\nu}^{(1)}\rangle$, since we could
always add a constant to $g_{\mu\nu}^{(1)}$ to make $\langle
g_{\mu\nu}^{(1)} \rangle$ anything we please.  The cancellation of the
non-gradient first-order terms in Eq.\ (\ref{combo31}) seems
accidental, but in Appendix \ref{alternative} we derive an expression
for $\langle\delta\theta\rangle$ that at first order only contains spatial
derivatives of $g_{\mu\nu}^{(1)}$, which do not change if we shift
$g_{\mu\nu}^{(1)}$ by a constant (or even a time-dependent renormalization).

\subsection{Second order}\label{secondorder}

The spatial metric $\gamma_{ij}$ is expanded up to second order
(neglecting first order vector and tensor perturbations) as\footnote{From 
now on spatial indices will be raised and lowered by the background metric
$\delta_{ij}$.}
\begin{equation}
\gamma_{i j} = \left(1-2\psi^{(1)}-\psi^{(2)}\right)\delta_{ij}
+D_{ij}\left(\chi^{(1)}+\frac{1}{2}\chi^{(2)}\right)  
+ \frac{1}{2}\left(\partial_i\chi_j^{(2)}+\partial_j\chi_i^{(2)}
+\chi_{ij}^{(2)}\right) .
\end{equation}
The functions $\psi^{(r)}$, $\chi^{(r)}$, $\chi_{i}^{(r)}$, and
$\chi_{ij}^{(r)}$ represent the $r$th-order perturbation of the
metric. Vector and tensor modes have been included here at second
order as they are dynamically generated by the non-linear evolution of
purely scalar perturbations \cite{MMB}.

It will turn out that $\psi^{(2)}({\bf x},\tau)$ is the only
second-order term for which we will require the explicit form.
For a matter-dominated Universe the metric perturbation 
$\psi^{(2)}({\bf x},\tau)$  is obtained similarly to  Ref. \cite{MMB}
\begin{equation}
\label{somp}
\psi^{(2)}({\bf x},\tau) = -\frac{50}{9} \varphi^2
- \frac{5\tau^2}{54} \varphi^{,k}\varphi_{,k}
+ \frac{\tau^4}{252} \left( \left(\nabla^2\varphi \right)^2
     -\frac{10}{3} \varphi^{,ki}\varphi_{,ki}
\right)  .
\end{equation}
The first term arises from a primordial epoch of inflation and can be
computed as follows. There exists a second-order extension of
the well-known gauge-invariant variable $\zeta$, 
the curvature perturbation on uniform density hypersurfaces. In terms of
${\mathcal H}=a'/a$, to first order it is given by
$\zeta^{(1)}=\psi^{(1)}+{\mathcal H} \delta^{(1)} \rho/\rho'$. To second
order, $\zeta=\zeta^{(1)}+(1/2)\zeta^{(2)}$, where $\zeta^{(2)}$
remains  {\it constant} on superhorizon scales in the case in which only
adiabatic perturbations are present.
In standard single-field inflation, $\zeta^{(2)}$ is generated during
inflation and its value is given by $\zeta^{(2)} \simeq -2\left(
\zeta^{(1)} \right)^2$ \cite{Acquavivaetal,BMR2}. Since
in the synchronous gauge and on superhorizon scales
$\zeta^{(1)}\simeq \psi^{(1)}=5\varphi/3$ and $\zeta^{(2)}\simeq
\psi^{(2)}$,
one readily concludes that the primordial contribution to the second-order
metric perturbation $\psi^{(2)}$ is given by the 
constant term $-50\varphi^2/9$ and it also propagates
to the coefficients of the term proportional to $\tau^2$.

In the second-order calculation care must be taken in defining the
spatial average.  Again, if ${\cal O}$ is a homogeneous quantity, then
simply $\langle {\cal O}^{(0)} \rangle ={\cal O}^{(0)}$. If $\cal O$
is already a second-order quantity, then we can take $\gamma=1$, both
in the numerator and in the denominator. But in the second-order
calculation of $\langle {\cal O}^{(1)} \rangle$ we must remember to 
include $\sqrt{\gamma}$. Using Eq.\ (\ref{fosd}), 
\begin{eqnarray}
\langle{\cal O}^{(1)} \rangle & = & 
\frac{\int d^3\!x \, \sqrt{\gamma(\tau,x^i)} \ 
{\cal O}^{(1)}(\tau,x^i)}{\int d^3 x\sqrt{\gamma(\tau,x^i)}} =
\frac{\int d^3\!x \, \left(1 -3\psi^{(1)}\right) 
{\cal O}^{(1)}(\tau,x^i)}{\int d^3\!x \, \, \left(1 -3\psi^{(1)}\right)}
\nonumber \\
& = & \frac{\int d^3\!x \,{\cal O}^{(1)}-3\int d^3\!x\,\psi^{(1)}\,
{\cal O}^{(1)}} {\int d^3\!x \ -3\int d^3\!x \, \psi^{(1)}}
= \langle{\cal O}^{(1)}\rangle_1 
- 3\langle\psi^{(1)}{\cal O}^{(1)}\rangle
+ 3\langle\psi^{(1)}\rangle_1
   \langle{\cal O}^{(1)}\rangle_1 ,
   \label{soa}
\end{eqnarray}
where we have introduced the notation $\langle \cdots \rangle_1$ for a
first-order term to denote
the spatial average with the factor $\sqrt{\gamma}=1$, {\it i.e.,}
$\langle {\cal O}^{(1)}\rangle_1(\tau)\equiv \int d^3\!x \, {\cal
O}^{(1)}(\tau,x^i)/\int d^3\!x $.

So the correct second-order averaging procedure gives
\begin{subequations}
\begin{eqnarray}
\langle{\cal O}^{(0)}\rangle & = & {\cal O}^{(0)} \\
\langle{\cal O}^{(1)}\rangle & = & 
\langle{\cal O}^{(1)}\rangle_1 
- 3\langle\psi^{(1)}{\cal O}^{(1)}\rangle
+ 3\langle\psi^{(1)}\rangle_1
    \langle{\cal O}^{(1)}\rangle_1
\label{avgone} \\
\langle{\cal O}^{(2)}  \rangle & = & 
\frac{\int d^3\!x \, {\cal O}^{(2)}(\tau,x^i)}{\int d^3\!x} .
\end{eqnarray}
\end{subequations}
As an illustration of the subtleties in averaging, if at first 
order $\langle g_{\mu\nu}^{(1)}\rangle_1 =0$, where the first-order 
averaging prescription is defined in Eq.\ (\ref{ffo}), at second 
order the averaging prescription
for a first-order quantity is defined by Eq.\ (\ref{avgone}), and 
$\langle g_{\mu\nu}^{(1)} \rangle$ need not vanish.

Now we perform a second-order expansion. We first expand the
energy-momentum tensor and the metric as
\begin{subequations}
\label{ourway}
\begin{eqnarray}
T_{\mu \nu} & = & T^{(0)}_{\mu \nu}+T^{(1)}_{\mu \nu}+T^{(2)}_{\mu \nu} 
\label{defT} \\
g_{\mu \nu} & = & g^{(0)}_{\mu \nu}+g^{(1)}_{\mu \nu}+g^{(2)}_{\mu \nu} ,  
\label{defg}
\end{eqnarray}
\end{subequations}
where again the superscript $(r)$ represents the $r$th-order
perturbation. Again we will take $T^{(0)}_{\mu \nu}$ and $g^{(0)}_{\mu \nu}$
to be homogeneous and isotropic, so $T^{(0)}_{\mu \nu} =
\langle T^{(0)}_{\mu \nu} \rangle$ and 
$g^{(0)}_{\mu\nu} = \langle g^{(0)}_{\mu \nu} \rangle$.   Note that this is 
{\em not} equivalent to the statement that $T^{(0)}_{\mu \nu} = 
\langle T_{\mu \nu} \rangle$ and 
$g^{(0)}_{\mu\nu} = \langle g_{\mu \nu} \rangle$. 

Expanding the Einstein tensor $G_{\mu\nu}$ to second order yields
several types of terms. The first type are the zeroth-order terms,
denoted $G_{\mu \nu}^{(0)}$, which only involve $g^{(0)}_{\mu \nu}$.
$G^{(0)}_{\mu \nu}$ will be homogeneous since $g^{(0)}_{\mu \nu}$ is
homogeneous by definition.  Then there are first-order terms, $G_{\mu
\nu}^{(1)}$,which involve a single power of $g^{(1)}_{\mu \nu}$,
possibly combined with $g^{(0)}_{\mu \nu}$.  There are two types of
second-order terms.  The first type, denoted by $G_{\mu \nu}^{(11)}$,
involves squares of $g^{(1)}_{\mu \nu}$. The second type of
second-order terms, denoted $G_{\mu \nu}^{(2)}$, involves
$g^{(2)}_{\mu \nu}$.

Now we consider the Einstein equations 
\begin{equation}
G_{\mu \nu}^{(0)} + G_{\mu \nu}^{(1)}+G_{\mu \nu}^{(11)}+ G_{\mu \nu}^{(2)} =
\kappa^2 \left( T_{\mu \nu}^{(0)} + T_{\mu \nu}^{(1)} +  T_{\mu \nu}^{(2)} 
\right) .
\label{pee}
\end{equation}
We proceed by averaging Eq.\ (\ref{pee}).  This yields
\begin{equation} 
G_{00}^{(0)}=\kappa^2\left\langle T_{00}\right\rangle
- \left\langle G_{00}^{(1)} 
+ G_{00}^{(11)}+G_{00}^{(2)} \right\rangle . \label{backEinstein}
\end{equation}
From Eq.\ (\ref{backEinstein}) we see that now $\kappa^{-2} \left\langle
G_{00}^{(1)} + G_{00}^{(11)}+G_{00}^{(2)} \right\rangle$
may be interpreted as an extra component to the stress-energy tensor.
To second order, the $0-0$ component of the perturbed FLRW model gives
\begin{equation}
3\left( \frac{a'}{a^2} \right)^2 = 
\kappa^2 \langle\rho\rangle - a^{-2}\left\langle G_{00}^{(1)} +
G_{00}^{(11)}+G_{00}^{(2)} \right\rangle .  \label{backEinstein00}
\end{equation}
Again we see that $\dot{a}/a=a'/a^2$ is {\em not} $\kappa^2 \rho^{(0)} /3$.

Now let us consider the second-order expression for the evolution of 
$\langle\rho\rangle$.  We again use the continuity equation,
Eq.\ (\ref{continuity}), and expand $\rho$ and $\theta$ as
\begin{eqnarray}
\rho   & = & \rho^{(0)}+\rho^{(1)}+\rho^{(2)}     \nonumber  \\ 
\theta & = &\theta^{(0)}+\theta^{(1)} + \theta^{(11)} + \theta^{(2)}  .      
\end{eqnarray}
The result is a bit more complicated than the first-order result:
\begin{equation}
\frac{1}{a}\left\langle\rho^{(0)\prime}+\rho^{(1)\prime}+\rho^{(2)\prime}
\right\rangle = - \left\langle \left(\theta^{(0)}+\theta^{(1)}
+ \theta^{(11)} + \theta^{(2)}\right)\left( \rho^{(0)} +\rho^{(1)}+\rho^{(2)}
 \right) \right\rangle      .        
\label{coneq}
\end{equation}
Using the fact that $\langle\rho^{(1)\prime}\rangle =
\langle\rho^{(1)}\rangle^\prime 
-3\langle\rho^{(1)}\rangle_1\langle\psi^{(1)\prime}\rangle_1
+3\langle\psi^{(1)\prime}\rho^{(1)}\rangle$, the left-hand side of 
Eq.\ (\ref{coneq}) becomes 
\begin{equation}
\frac{1}{a}\left\langle\rho^{(0)\prime}+\rho^{(1)\prime}+\rho^{(2)\prime}
\right\rangle 
= \frac{1}{a}\langle\rho\rangle' 
- \frac{3}{a}  \langle\psi^{(1)\prime}\rangle_1 \langle\rho^{(1)}\rangle_1
+ \frac{3}{a}\langle\psi^{(1)\prime}\rho^{(1)}\rangle .
\label{lefths}
\end{equation}
We may also express the right-hand side of Eq.\ (\ref{coneq}) as
\begin{equation}
 - \left\langle \left(\theta^{(0)}+\theta^{(1)}
+ \theta^{(11)} + \theta^{(2)}\right)\left( \rho^{(0)} +\rho^{(1)}+\rho^{(2)}
 \right) \right\rangle  
 =
- \langle \theta \rangle\langle \rho\rangle 
+ \langle \theta^{(1)}\rangle_1 \langle\rho^{(1)}\rangle_1 
- \langle\rho^{(1)}\theta^{(1)}\rangle .
\label{righths}
\end{equation}
Equating Eqs.\ (\ref{lefths}) and (\ref{righths}) we obtain
\begin{equation}
\frac{1}{a}\langle\rho\rangle' 
- \frac{3}{a} \langle\psi^{(1)\prime}\rangle_1 \langle\rho^{(1)}\rangle_1 
+ \frac{3}{a}\langle\psi^{(1)\prime}\rho^{(1)}\rangle
=
- \langle \theta \rangle\langle \rho\rangle 
+ \langle \theta^{(1)}\rangle_1 \langle\rho^{(1)}\rangle_1 
- \langle\rho^{(1)}\theta^{(1)}\rangle .
\label{almost}
\end{equation}
Since to first order a simple calculation yields 
$\theta^{(1)}=-3\psi^{(1)^\prime}/a$, we are left with
\begin{equation}
 \frac{1}{a} \frac{\langle\rho\rangle'}{\langle\rho\rangle} 
= - \langle \theta \rangle  .
\label{continuityAveraged} 
\end{equation}
This tells us that as in a first-order perturbed universe, 
in a second-order perturbed Universe the average
matter density is diluted with expansion according to
$\langle\rho\rangle^{-1} d\langle\rho\rangle/dt = -\langle \theta
\rangle$, which in general is {\it not} equivalent to
$\langle\rho\rangle^{-1} d\langle\rho\rangle/dt = -3\dot{a}/a$.  Note
that it was crucial to define the averages with the $\sqrt{\gamma}$
factor; otherwise we would not have discovered the right quantity 
to describe the scaling of $\langle \rho \rangle$.

Again, the physical quantity of interest is $\langle\theta\rangle$, given by
\begin{equation}
\langle\theta\rangle = \theta^{(0)} + \langle\theta^{(1)}\rangle
+ \langle\theta^{(11)}\rangle + \langle\theta^{(2)}\rangle
= 3\frac{a'}{a^2} + \langle\theta^{(1)}\rangle
+ \langle\theta^{(11)}\rangle + \langle\theta^{(2)}\rangle .
\end{equation}
We emphasize again that 
$H\equiv\sqrt{\kappa^2\langle\rho\rangle/3}$ (and {\it not} $a'/a^2$). 

The goal of this paper is to calculate $\langle\delta\theta\rangle/3H$.  
To do so, we can express Eq.\ (\ref{backEinstein00}) in the form
\begin{equation}
\frac{a'}{a^2} = \left( \frac{\kappa^2 \langle\rho\rangle}{3} 
- \frac{\left\langle G_{00}^{(1)} + G_{00}^{(11)}+G_{00}^{(2)} \right\rangle}
{3a^2} \right)^{1/2} \simeq H 
\left(1-\frac{\left\langle G_{00}^{(1)}+G_{00}^{(11)}+G_{00}^{(2)}\right\rangle}
{6a^2H^2}- \frac{\left\langle G^{(1)}_{00}\right\rangle^2}{72a^4H^4}\right) ,
\label{combo2}
\end{equation}
where of course the second equality holds if the corrections are small.   
Using  Eq.\ (\ref{combo2}), we find
\begin{equation}
\frac{\langle\delta\theta\rangle}{3H} = \frac{\left\langle\theta^{(1)}
+ \theta^{(11)} + \theta^{(2)}\right\rangle}{3H} 
- \frac{\left\langle G_{00}^{(1)} + G_{00}^{(11)}+G_{00}^{(2)}\right\rangle} 
{6a^2H^2} - \frac{\left\langle G^{(1)}_{00}\right\rangle^2}{72a^4H^4} .
\label{combo3}
\end{equation}

It will turn out that $g_{\mu\nu}^{(2)}$ will appear only as a spatial
gradient in the final expression for $\langle\delta\theta\rangle$.
This means that now the physical result is insensitive to the choice of
the normalization of $\langle g_{\mu\nu}^{(2)}\rangle$, since we could
always add a term that is spatially constant.  The cancellation of the
non-gradient second-order terms in Eq.\ (\ref{combo3}) seems
accidental, but again, in Appendix \ref{alternative} we derive an expression
for $\langle\delta\theta\rangle$ that explicitly only contains spatial
derivatives of $g_{\mu\nu}^{(2)}$, which do not change if we shift
$g_{\mu\nu}^{(2)}$ by a term that is spatially constant (or even a
time-dependent normalization).

\section{Computation of $\langle\delta\theta\rangle/3H$ in the synchronous 
gauge  \label{thetasynchronous}} 

The relevant quantity which we want to calculate is
$\langle\delta\theta\rangle/3H$. The simplest way to proceed is to
perform the computation directly in the synchronous gauge. The
synchronous coordinates are more physical for our purposes.  If the
calculation is performed in the Poisson gauge it is necessary to
perform a complex calculation (as we do in Appendix
\ref{thetapoisson}) to express the result in synchronous coordinates
(see also Refs.\ \cite{Rasanen} and
\cite{geshzjiani}).

Moreover, in general for second-order calculations, the synchronous 
gauge turns out to be very convenient, since here the scalar perturbations 
do not have nonlocal terms that appear in other gauges such as the Poisson 
gauge.  Also, for matter $u^{\mu}$ in this gauge takes the trivial form
$u^{\mu}=a^{-1}(\tau)(1,\vec{0})$, so the only terms that have to be
computed are the Christoffel symbols in the covariant derivative
\begin{equation}
\theta=D_{\mu}u^{\mu} 
= \partial_{\mu}u^{\mu}+\Gamma_{\mu \alpha}^{\alpha} u^{\mu}
= 3\frac{a'}{a^2} 
+ \frac{\delta\Gamma^0_{0 0}+ \delta\Gamma^i_{0 i}}{a}  .
\end{equation}
Moreover, in the synchronous gauge, to second order $\delta\Gamma_0^{0
0}=0$ \cite{Acquavivaetal}. Finally, the computation of the
perturbation of $\theta$ with respect to its background value consists
of finding just the trace of $\Gamma^i_{0 j}$.
From \cite{Acquavivaetal}, 
\begin{eqnarray}
\frac{\left\langle\theta^{(1)} + \theta^{(11)} + \theta^{(2)}\right\rangle}{3H}
 & = & \frac{1}{3H}\left\langle\delta\Gamma^i_{0 i}\right\rangle = 
 \frac{1}{3Ha}\left\langle\delta^{(1)}
\Gamma^i_{0 i}+\delta^{(2)}\Gamma^i_{0 i}\right\rangle \nonumber \\
& = & \frac{1}{aH} \left\langle-\psi^{(1)\prime} - \frac{1}{2} \psi^{(2)\prime}
- 2\psi^{(1)}\psi^{(1)\prime}
+ \frac{1}{18} \nabla^2 \chi^{(1)}\nabla^2\chi^{(1)\prime} 
-\frac{1}{6}\chi^{(1), kj}\
\chi_{\ \ \ ,jk}^{(1)\prime} \right\rangle \nonumber \\
& = & \frac{1}{a^2H^2}\left\langle -\frac{2}{9} \nabla^2\varphi
- \frac{20}{27} \varphi\nabla^2\varphi
+ \frac{5}{27}\varphi_{,i}\varphi^{,i} 
- \frac{\tau^2}{63}\left(\nabla^2\varphi\right)^2
- \frac{4\tau^2}{189}\varphi^{,ij}\varphi_{,ij} \right\rangle.
\label{fundamentalsyn}
\end{eqnarray}
In the last equation we expressed the result in terms of the initial
perturbation $\varphi$, using the time evolution of the relevant
perturbation variables in the synchronous gauge given in Eqs.\
(\ref{fomp}) and (\ref{somp}).

The simplest way to find $G_{00}$ is to realize that
$a^{-2}\left(G_{00}^{(1)}+G_{00}^{(11)}+G_{00}^{(2)}\right)=\kappa^2
\left(\rho^{(1)}+\rho^{(2)}\right)$, and use  the explicit solutions
for $\rho^{(1)}$ and $\rho^{(2)}$ from Ref.\ \cite{MMB}, but using for
$\psi^{(2)}$ the expression augmented with the $\varphi^2$ as in 
Eq.\ (\ref{somp}).    In the matter-dominated universe,
\begin{eqnarray}
- \frac{\left\langle G_{00}^{(1)} +G_{00}^{(11)}+G_{00}^{(2)}\right\rangle}
            {6a^2H^2} & = &
  \frac{1}{a^2H^2} \left\langle
 -\frac{1}{3}\nabla^2 \varphi  
 -\frac{10}{27} \varphi \nabla^2 \varphi
 +\frac{55}{54} \varphi_{, i} \varphi^{, i}
 -\frac{5\tau^2}{126} \left(\nabla^2 \varphi\right)^2
 -\frac{\tau^2}{63} \varphi^{, ij}\varphi_{, ij} \right\rangle \nonumber \\
 -\frac{\left\langle G_{00}^{(1)}\right\rangle^2}{72a^4H^4} & = & 
 - \frac{1}{a^2H^2}
  \frac{\tau^2}{72}\langle\nabla^2 \varphi\rangle\langle\nabla^2 
  \varphi\rangle  .
\label{fundamentalsyng}
\end{eqnarray}
  
The value of $\delta\theta/3H$ is found by summing Eqs.\ 
(\ref{fundamentalsyn}) and (\ref{fundamentalsyng}). 

We must now perform the appropriate spatial average. In $\delta\theta$
there is only one first-order term; the rest of the terms are second order.
The first-order term must be averaged using the procedure of Eq.\
(\ref{soa}), while the other terms are averaged using
$\sqrt{\gamma}=1$.  The result is
\begin{eqnarray}
\frac{\langle\delta\theta\rangle}{3H} & = & \frac{1}{a^2H^2} \left[
- \frac{5}{9} \left\langle\nabla^2\varphi\right\rangle_1
+ \frac{5}{3} \left( \left\langle\varphi\nabla^2\varphi\right\rangle
+ \frac{13}{18}\left\langle\varphi_{,i}\varphi^{,i}\right\rangle \right)
+ \frac{\tau^2}{27}\left(\left\langle\left(\nabla^2\varphi\right)^2
\right\rangle -\left\langle\varphi^{,ij}\varphi_{,ij}\right\rangle\right)
\right . \nonumber \\
& & \left. - \frac{25}{9}\left\langle\varphi\right\rangle_1
\left\langle\nabla^2\varphi\right\rangle_1
- \frac{23\tau^2}{216}\left\langle\nabla^2\varphi\right\rangle_1
\left\langle\nabla^2\varphi\right\rangle_1 \right]  .   
\label{risultatoWithBt}
\end{eqnarray}

\subsection{Evaluation of $\langle\delta\theta\rangle$ in terms of the 
matter power spectrum \label{k^4}}

We now proceed to express the averages in terms of the matter power
spectrum.  The procedure is to fix a spherical domain of radius $R$
with volume $V(R)$.  From Eq.\ (\ref{average}) and the definition of
$\langle \cdots \rangle_1$, all of the averages in Eq.\
(\ref{risultatoWithBt}) involve integrals of the form
\begin{equation}
\langle\cdots \rangle \equiv \frac{1}{V(R)} \int_{V(R)} (\cdots)  \ d^3\!x .
\end{equation}
For calculational convenience we will employ a Gaussian window function and
assume $V(R)$ is a spherically symmetric volume with volume 
element $dV=4\pi r^2 \exp(-r^2/2R^2)dr$ and volume $V(R)=(2\pi)^{3/2}R^3$.  
The Fourier transform of the window functions is
\begin{equation}
W(kR) = V^{-1}(R) \int d^3\!x\ e^{-r^2/2R^2} \exp(i \vec{k}\cdot \vec{x})
= e^{-k^2R^2/2}.
\end{equation}
Of course as $kR\rightarrow0$, $W(kR)\rightarrow 1$.

We wish to evaluate the typical expected value of $\theta$ averaged
over this sphere. By ``typical expected value'' we mean the ensemble
average.  The metric fluctuation $\varphi$ is treated as a Gaussian
variable with zero mean (of which we know the $N$-point correlation 
functions) that
takes random values over different ``realizations'' of volumes $V(R)$.
In other words, we calculate the typical value of a quantity for a
region of radius $R$ as the statistical mean over many different
similar regions. We will indicate this statistical average with a bar:
$\overline{\langle\cdots\rangle}$.

We will express $\varphi$ and its derivatives in terms of a Fourier
integral, so
\begin{equation}
\varphi = \int\frac{d^3\!k}{(2\pi)^3}
\ \varphi_{\vec{k}} \ e^{i\vec{k}\cdot\vec{x}} , \quad
\varphi_{,i} = \int\frac{d^3\!k}{(2\pi)^3} \ ik_i
\ \varphi_{\vec{k}} \ e^{i\vec{k}\cdot\vec{x}} , \quad 
\nabla^2\varphi = - \int\frac{d^3\!k}{(2\pi)^3} \ k^2
\ \varphi_{\vec{k}} \ e^{i\vec{k}\cdot\vec{x}} , \quad \textit{etc.}
\end{equation}
The Fourier components $\varphi_{\vec{k}}$ satisfy 
\begin{subequations}
\begin{eqnarray}
\overline{\varphi_{\vec k}} & = & 0 
\label{onepoint} \\
\overline{\varphi_{\vec{k}_1} \varphi_{\vec{k}_2}}
 & = & (2 \pi)^3 \delta^{(3)}(\vec{k_1}+\vec{k_2}) \ 
 P_{\varphi}(k_1)\label{twopoint}\\
\overline{\varphi_{\vec{k}_1}\varphi_{\vec{k}_2}
\varphi_{\vec{k}_3} \varphi_{\vec{k}_4}} 
& = & (2\pi)^6
\left\{\delta^{(3)}(\vec{k}_1+\vec{k}_2) \delta^{(3)}(\vec{k}_3+\vec{k}_4)
\ P_{\varphi}(k_1)P_{\varphi}(k_3)  \right. \nonumber \\
& & \left.+\left[ \delta^{(3)}(\vec{k}_1+\vec{k}_3)
\delta^{(3)}(\vec{k}_2+\vec{k}_4)
+\delta^{(3)}(\vec{k}_1+\vec{k}_4)\delta^{(3)}(\vec{k}_2+\vec{k}_3) \right]
\ P_{\varphi}(k_1)P_{\varphi}(k_2)    
\right\}  \label{fourpoint} ,
\end{eqnarray}
\end{subequations}
where $P_\varphi(k) = \left| \varphi_{\vec{k}}\right|^2$.  From Eq.\
(\ref{nabla2varphi}) we can express $P_\varphi(k)$ in terms of the
matter power spectrum as
\begin{equation}
P_{\varphi}(k) \equiv \frac{9\pi^2}{2} \ a^4 H^4 \ \frac{\Delta^2(k,a)}
{k^7},
\label{pphipdelta}
\end{equation}
where $\Delta^2(k,a)$ is the (dimensionless) power spectrum of the
matter density fluctuations.

Let us first consider $\overline{\left\langle\varphi\right\rangle_1}$ and
$\overline{\left\langle\nabla^2\varphi\right\rangle_1}$. Clearly from
Eq.\ (\ref{onepoint}), $\overline{\left\langle\varphi\right\rangle_1}=0$
and  $\overline{\left\langle \nabla^2 \varphi
\right\rangle_1}=0$.  However this does not imply that 
$\left\langle\varphi\right\rangle_1=0$ or $\left\langle\nabla^2
\varphi \right\rangle_1=0$ over any individual volume of radius $R$.
The question is the magnitude of typical departures from the mean
values, which corresponds to the statistical variance of our
quantities.  As we will show, it is intuitively clear that if the
radius $R$ is big enough, this variance will go to zero.  So the
effect of variance could be important.  We will return to the
calculation of the variances of the different terms after completing
the calculation of the mean values.

Next, consider $\overline{\left\langle \varphi \nabla^2 \varphi
\right\rangle}$. Passing to Fourier space, we have
\begin{eqnarray}
\overline{\left\langle \varphi \nabla^2 \varphi \right\rangle}
& = &-\int_{V(R)}\frac{d^3\!x}{V(R)} \frac{d^3\!k_1}{(2\pi)^3}
\frac{d^3\!k_2}{(2 \pi)^3}\left( \frac{k_1^2 + k_2^2}{2}  \right) 
\overline{\varphi_{\vec{k_1}} \varphi_{\vec{k_2}}} \ 
\exp\left[i(\vec{k}_1+\vec{k}_2)\cdot \vec{x}\right]  \nonumber  \\
& = & -  \int \frac{d^3\!k_1}{(2\pi)^3} \frac{d^3\!k_2}{(2 \pi)^3}
\left( \frac{k_1^2 + k_2^2}{2}  \right) 
\overline{\varphi_{\vec{k}_1} \varphi_{\vec{k}_2}} \
W\left(\left|\vec{k_1} +\vec{k_2}\right| R\right) .
\end{eqnarray}
Making use of Eq.\ (\ref{twopoint}), we find 
\begin{equation}
\overline{\left\langle \varphi \nabla^2 \varphi \right\rangle} = 
- \int \frac{d^3\!k}{(2 \pi)^3}\, k^2 P_{\varphi}(k) .
\end{equation}
Using Eq.\ (\ref{pphipdelta}), we can express $\overline{\left\langle
\varphi \nabla^2 \varphi \right\rangle}$ in its final form.  In the
same manner we find the means for the other terms.  The result is
\begin{subequations}
\begin{eqnarray}
\overline{\left\langle\varphi\right\rangle_1} & = & 0 \\
\overline{\left\langle\nabla^2\varphi\right\rangle_1} & = & 0 \\
\overline{\left\langle \varphi \nabla^2 \varphi \right\rangle} & = &
-\frac{9}{4} a^4 H^4 \int_0^{\infty} \frac{dk}{k^3} \Delta^2(k,a) \\
\overline{\left\langle\varphi_{,i}\varphi^{,i}\right\rangle} & = &
+\frac{9}{4} a^4 H^4 \int_0^{\infty} \frac{dk}{k^3} \Delta^2(k,a) \\
\overline{\left\langle\left(\nabla^2\varphi\right)^2\right\rangle} & = &
+\frac{9}{4} a^4 H^4 \int_0^{\infty} \frac{dk}{k} \Delta^2(k,a)\\
\overline{\left\langle\varphi^{,ij}\varphi_{,ij}\right\rangle} & = & 
+\frac{9}{4} a^4 H^4 \int_0^{\infty} \frac{dk}{k} \Delta^2(k,a) \\
\overline{\langle\varphi\rangle_1\langle\nabla^2\varphi\rangle_1} & = &
-\frac{9}{4} a^4 H^4 \int_0^{\infty} \frac{dk}{k^3} \Delta^2(k,a) \, W^2(kR) 
\label{ggg} \\
\overline{\langle\nabla^2\varphi\rangle_1\langle\nabla^2\varphi\rangle_1} & = &
+\frac{9}{4} a^4 H^4 \int_0^{\infty} \frac{dk}{k} \Delta^2(k,a) \, W^2(kR) .
\label{hhh}
\end{eqnarray}
\end{subequations}
Of course in the limit $R\rightarrow\infty$, Eqs.\ (\ref{ggg}) and 
(\ref{hhh}) vanish.  Therefore, the entire second-order calculation gives
\begin{equation}
\frac{\overline{\left\langle\delta\theta\right\rangle}}{3H} 
= -\frac{25}{24} a^2H^2  \int_0^{\infty} \frac{dk}{k^3} \Delta^2(k,a)
+ \frac{25}{4}a^2H^2 \int_0^{\infty} \frac{dk}{k^3} \Delta^2(k,a)W^2(kR)
- \frac{23}{96} \int_0^{\infty} \frac{dk}{k} \Delta^2(k,a)W^2(kR) ,
\label{thisisit}
\end{equation}
where we have used $\tau = 2/aH$ appropriate for a matter-dominated
Universe.  We will give numerical results in the next section.

Now for the variances of selected terms. The variance is defined as
\begin{equation}
\textrm{Var}\left[\left\langle\cdots\right\rangle\right] = 
\overline{ \left( \left\langle\cdots\right\rangle - 
\overline{\left\langle\cdots\right\rangle} \right)^2 } .
\end{equation}
For instance,
\begin{subequations}
\begin{eqnarray}
\textrm{Var}\left[\left\langle\varphi\right\rangle_1\right] & = &  
\frac{9}{4} a^4 H^4 \int_0^\infty \frac{dk}{k^5} \Delta^2(k,a) \ W^2(kR) 
\label{varphi}  \\
\textrm{Var}\left[\left\langle\nabla^2\varphi\right\rangle_1\right] & = &  
\frac{9}{4} a^4 H^4 \int_0^\infty \frac{dk}{k} \Delta^2(k,a) \ W^2(kR)  .
\label{varlinear}
\end{eqnarray}
\end{subequations}

The variances of other terms are more complicated, but straightforward to 
derive. For instance,
\begin{equation}
\textrm{Var} \left[ \left\langle\varphi\nabla^2\varphi\right\rangle
+ \frac{13}{18}\left\langle\varphi_{,i}\varphi^{,i}\right\rangle\right]
=\frac{1}{(2 \pi)^6}  \int d^3\!k_1  d^3\!k_2
\frac{1}{2}\left[ k_1^2+k_2^2+\frac{13}{9}\vec{k}_1\cdot\vec{k}_2  \right]^2
W^2(|\vec{k_1}+\vec{k_2}| R ) \ P_{\varphi}(k_1) P_{\varphi}(k_2)  .
\label{twopointvar}
\end{equation}
The angular integrals can be expressed in terms of a filter function, 
defined in general as
\begin{equation}
J^{(l)}(k_1,k_2,R) \equiv \int_{-1}^{1}d\mu\,\mu^l \ W^2
\left(\sqrt{k_1^2+k_2^2+2  k_1 k_2 \mu}\, R\right) .
\end{equation}
For a Gaussian window function the filter expression can be expressed in
terms of incomplete $\Gamma$ functions as
\begin{eqnarray}
J^{(l)}(k_1,k_2,R) & = & \int_{-1}^1d\mu \ 
           \mu^l \ e^{-(k_1^2+k_2^2+2\mu k_1k_2)R^2} \nonumber \\
& = & \frac{\exp\left[-\left(k_1^2 + k_2^2 \right)R^2\right]
\left[\Gamma(l+1,-\,2R^2 {k_1} {k_2}) - \Gamma(l+1, 2R^2{k_1}{k_2})\right]}
{{2^{l+1}R^{2(l+1)}{k_1}}^{l+1}{{k_2}}^{l+1} }  .
\end{eqnarray}
We will make use of the fact that for $k_1R\rightarrow 0$ and 
$k_2R\rightarrow 0$, $J^{(0)}(k_1,k_2,R)\rightarrow 2$.

Using the Gaussian filter, Eq.\ (\ref{twopointvar}) becomes 
\begin{eqnarray}
\textrm{Var} \left[ \left\langle\varphi\nabla^2\varphi\right\rangle
+ \frac{13}{18}\left\langle\varphi_{,i}\varphi^{,i}\right\rangle\right]
& = &\left( \frac{9}{8}  a^4 H^4 \right)^2
\int_0^{\infty} \frac{d k_1}{k_1^3} \Delta^2(k_1,a)
\int_0^{\infty} \frac{d k_2}{k_2^3} \Delta^2(k_2,a) 
\left[ \left(\frac{k_1^2}{k_2^2}+\frac{k_2^2}{k_1^2}+2\right)
J^{(0)}(k_1,k_2,R)\right.\nonumber\\
& & \left. +\frac{26}{9}\left( \frac{k_1}{k_2}+\frac{k_2}{k_1}\right) 
J^{(1)}(k_1,k_2,R)+\frac{169}{81}J^{(2)}(k_1,k_2,R) \right] .
\label{phi2k2}
\end{eqnarray}
It is interesting that the integral is not well behaved in the infrared.  As
discussed in the next section, for a Harrison--Zel'dovich spectrum 
$\Delta^2(k)\propto k^4$, so the term proportional to $J^{(0)}$ has an 
infrared divergence:
\begin{eqnarray}
\textrm{Var}\left[\left\langle\varphi\nabla^2\varphi\right\rangle\right]_\infty
& = & \left( \frac{9}{8}  a^4 H^4 \right)^2
\int_0^{\infty} \frac{d k_1}{k_1^3} \Delta^2(k_1,a)
\int_0^{\infty} \frac{d k_2}{k_2^3} \Delta^2(k_2,a) 
\left(\frac{k_1^2}{k_2^2}+\frac{k_2^2}{k_1^2}\right) J^{(0)}(k_1,k_2,R)
\nonumber \\
& = & \left( \frac{9}{8}  a^4 H^4 \right)^2 4
\int\frac{dk_2}{k_2}\Delta^2(k_2,a) \int\frac{dk_1}{k_1^5}\Delta^2(k_1,a) .
\end{eqnarray}
We will discuss the importance of this term below.

Now we turn to the variances of 
$\left\langle\varphi\right\rangle_1
\left\langle\nabla^2\varphi\right\rangle_1$ and 
$\left\langle\nabla^2\varphi\right\rangle
\left\langle\nabla^2\varphi\right\rangle$.  
They are given by
\begin{subequations}
\label{uytr}
\begin{eqnarray}
\textrm{Var}\left[\left\langle\varphi\right\rangle_1
\left\langle\nabla^2\varphi\right\rangle_1\right] & = & 
\left(\frac{9}{4}a^4H^4\right)^2
\left[\int \frac{dk_1}{k_1}\Delta^2(k_1,a)\,W^2(k_1R)
\int \frac{dk_2}{k_2^5} \Delta^2(k_2,a)W^2(k_2R) \right. \nonumber \\ 
& & \left. +  \left( \int\frac{dk}{k^3}\Delta^2(k,a)W^2(kR)\right)^2 \right] 
\label{another} \\
\textrm{Var}\left[ \left\langle\nabla^2\varphi\right\rangle
\left\langle\nabla^2\varphi\right\rangle \right] & = & 
2\left(\frac{9}{4}a^4H^4\right)^2
\left( \int\frac{dk}{k}\Delta^2(k,a)W^2(kR)\right)^2 .
\end{eqnarray}
\end{subequations}
Note that Eq.\ (\ref{another}) also has an infrared divergence
\begin{equation}
\textrm{Var}\left[\left\langle\varphi\right\rangle_1
\left\langle\nabla^2\varphi\right\rangle_1\right]_\infty = 
\left( \frac{9}{8}  a^4 H^4 \right)^2 4
\int\frac{dk_2}{k_2}\Delta^2(k_2,a) \int\frac{dk_1}{k_1^5}\Delta^2(k_1,a) .
\end{equation}

Of course we will be interested in the variance of the total expression Eq.\
(\ref{risultatoWithBt}), not the individual terms.  Of particular interest is
the cross term of the infrared-singular parts.  This will be the only one for
which we will include the cross terms.  The infrared-singular pieces
appear in Eq.\ (\ref{risultatoWithBt}) proportional to 
$\langle\varphi\nabla^2\varphi\rangle
-3\langle\varphi\rangle_1\langle\nabla^2\varphi\rangle_1$.  The infrared 
part of the variance of this term is
\begin{equation}
\textrm{Var}[\langle\varphi\nabla^2\varphi\rangle
-3\langle\varphi\rangle_1\langle\nabla^2\varphi\rangle_1]_\infty =
\left(\frac{9}{4}  a^4 H^4 \right)^2 4
\int\frac{dk_2}{k_2}\Delta^2(k_2,a) \int\frac{dk_1}{k_1^5}\Delta^2(k_1,a) .
\label{blowup}
\end{equation}

It is straightforward to obtain the variance for the second-order, 
four-derivative terms. (The sum of these terms has zero mean.) It is given by
\begin{eqnarray}
\textrm{Var} \left[ \left\langle\left(\nabla^2\varphi\right)^2\right\rangle
- \left\langle\varphi_{,ij}\varphi^{,ij}\right\rangle\right] & = & 
 \left(\frac{9}{4}  a^4 H^4 \right)^2
\int_0^{\infty} \frac{d k_1}{k_1} \Delta^2(k_1,a)
\int_0^{\infty} \frac{d k_2}{k_2} \Delta^2(k_2,a)
\left[ J^{(0)}(k_1,k_2,R) -2  J^{(2)}(k_1,k_2,R) \right. \nonumber \\
& & \left.+J^{(4)}(k_1,k_2,R) \right]  .
\label{phi2k4}
\end{eqnarray}

There is another potential contribution that will result in contributions to the
variance similar to the terms we have found.  Suppose we expand $\delta\theta$
to {\em third} order in perturbation theory.  We can express $\delta\theta$ in
the general form
\begin{equation}
\left\langle\delta\theta\right\rangle = \left\langle {\mathcal A}\varphi 
+ {\mathcal B}\varphi^2 + {\mathcal C}\varphi^3 \right\rangle ,
\end{equation}  
where ${\mathcal A}$, ${\mathcal B}$, and ${\mathcal C}$ are operators which
contain derivatives. Then the variance of $\delta\theta$ will contains terms 
like $\langle{\mathcal A}\varphi\rangle^2$, $\langle{\mathcal B}
\varphi^2\rangle^2$, and 
$\langle{\mathcal A}\varphi\rangle\langle{\mathcal C}\varphi^3\rangle$. 
The first term is the usual cosmic
variance term; it is well behaved in the infrared.  The second term is singular
in the infrared; it is given in Eq.\ (\ref{blowup}).  The third term 
will be present, in principle it also will have an infrared singular part, 
and its value requires a relativistic third-order perturbation calculation. 
However there is no reason for the infrared singular part of the  
${\mathcal A}{\mathcal C}$ term to cancel exactly the infrared singular part of
the ${\mathcal B}^2$ term.

Finally we remark on the ultraviolet behavior of the corrections to the
expansion rate.  The second-order result for the mean, Eq.\ (\ref{thisisit}),
should be well behaved in the ultraviolet.  As discussed in the next section,
in the linear regime $\Delta^2(k,a)$ increases logarithmically with $k$ for a
Harrison-Zel'dovich spectrum, so the first term should be well behaved in the
ultraviolet.  The ultraviolet behavior of the last two terms are regulated by
the filter function $W^2(kR)$.  The contributions to the variance of the terms
we have calculated, Eqs.\ (\ref{varlinear}), (\ref{phi2k2}), (\ref{uytr}), and 
(\ref{phi2k4}),  all involve filter functions that regulate the ultraviolet 
behavior.  However, we expect there to be terms that do not involve filter 
functions.  For instance, if one performs the relativistic third-order 
calculation, one expects to find contributions to $\langle\delta\theta\rangle$ 
from terms like $\langle(\nabla^2\varphi)^3\rangle$.  
The variance would then include evaluation of terms like 
$\langle\nabla^2\varphi\rangle\langle(\nabla^2\varphi)^3\rangle$ 
(an example of the aforementioned ${\mathcal A}{\mathcal C}$ terms). These 
terms would include parts with a momentum integration unregulated by
a filter function.  If there are third-order terms with large numbers of
derivatives bringing down large powers of momentum, then the variance might be
sensitive to the ultraviolet behavior.

In the next section we show the numerical results for the mean and variance of
the corrections to the expansion rate.

\section{Numerical results}\label{numbercrunch}

In this section we present the numerical results for
$\langle\delta\theta\rangle/3H$ for a matter-dominated Universe.  We
will give the mean values, as well as the variances.

For both the mean and the variances we express the power spectrum
$\Delta^2(k,a)$ in terms of the transfer function $T^2(k)$.  For a
Harrison--Zel'dovich spectrum, the power spectrum is
\begin{equation}
\Delta^2(k,a) = A^2 \left(\frac{k}{aH}\right)^4 T^2(k) ,
\label{powerspectrum}
\end{equation}
where $A$ is the dimensionless amplitude, $A=1.9\times10^{-5}$.   
We will discuss the
implications of other spectra.  For our purposes the Bardeen, Bond,
Kaiser, Szalay (BBKS) transfer
function \cite{BBKS} will be adequate.  The BBKS transfer function may
be expressed in terms of a dimensionless parameter
\begin{equation}
q\equiv\frac{k}{\Gamma h \textrm{Mpc}^{-1}},
\end{equation}
where $\Gamma$ is the shape factor, defined for a flat universe in
terms of the baryon fraction $\Omega_B$ and the total value of
$\Omega_0$ as $\Gamma=\Omega_0 h
\exp(-\Omega_B-\sqrt{2}h\Omega_B/\Omega_0)$.  In terms of $q$,
\begin{equation}
T(q) = \frac{\ln\left(1+2.34q\right)}{2.34q} \left[
1+ 3.89q + (16.1q)^2 + (5.84q)^3 +(6.71q)^4\right]^{-1/4} .
\end{equation}
Of course at small $q$, $T^2(q)\rightarrow 1$, while at large $q$,
$T^2(q)\rightarrow q^{-4}\ln^2\!q$.

Also, in all expressions we make use of the fact that in a
matter-dominated Universe $H^2(a)=H_0^2a_0^3/a^3$, where $a_0$ is the
present value of the scale factor.

Consider the mean, given by Eq.\ (\ref{thisisit}):
\begin{eqnarray}
\frac{\overline{\left\langle\delta\theta\right\rangle}}{3H} 
& = & -\frac{25}{24}
\frac{a}{a_0}A^2 \left(\frac{h\, \textrm{Mpc}^{-1}}{H_0}\right)^2\Gamma^2
\int_0^\infty dq \ q \ T^2(q) 
+\frac{25}{4}
\frac{a}{a_0}A^2 \left(\frac{h\, \textrm{Mpc}^{-1}}{H_0}\right)^2\Gamma^2
\int_0^\infty dq \ q \ T^2(q) \, W^2(r\Gamma q) \nonumber \\ & & 
-\frac{23}{96} \left(\frac{a}{a_0}\right)^2
A^2 \left(\frac{h\, \textrm{Mpc}^{-1}}{H_0}\right)^4\Gamma^4
\int_0^\infty dq \ q^3 \ T^2(q) \, W^2(r\Gamma q) ,
\label{edsmean}
\end{eqnarray}
where $r$ is the dimensionless size of the region,
$r\equiv R/h^{-1}\textrm{Mpc}$.  We will present results for $R=H_0^{-1}$ (so
$r=3000$) and the last two terms are negligible.
While we have indicated the range of integration from $q=0$ to
$q=\infty$, in reality to employ Eq.\ (\ref{powerspectrum}) 
there is a cutoff on the
maximum value of the integral.  The ultraviolet cutoff arises because
density perturbations become nonlinear.  The mean value is very
insensitive to the ultraviolet cutoff. The integral for the mean value
receives most of the contribution in the decade between $10^{-1}
\leq q \leq 1$.   

\begin{figure}
\includegraphics[width=0.9\textwidth]{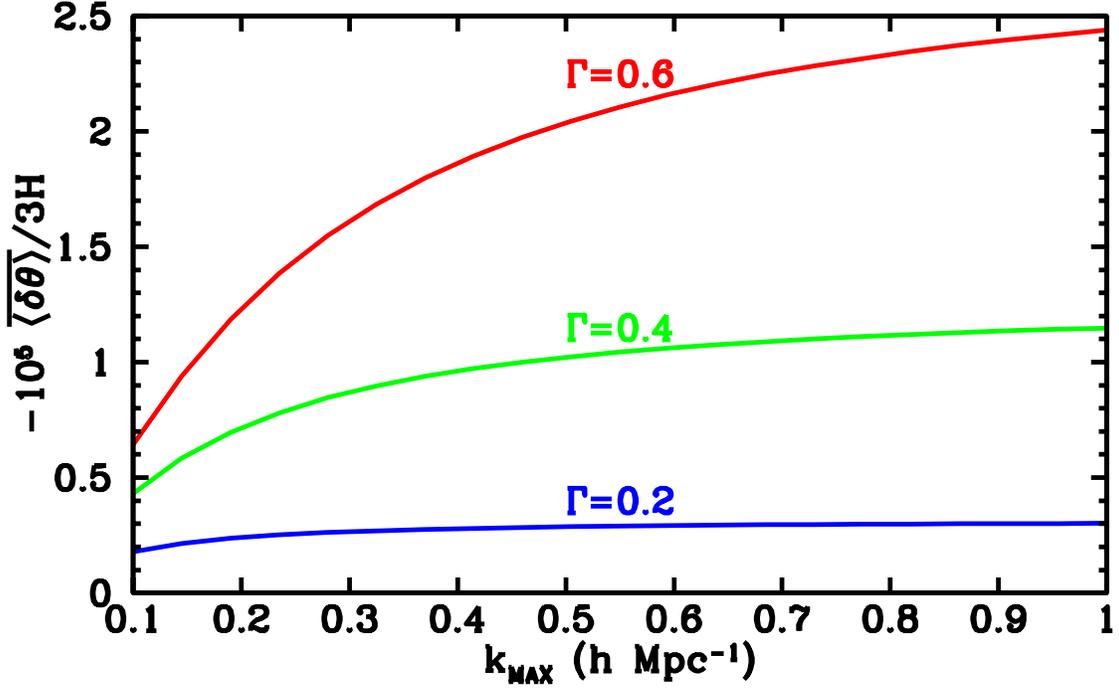}
\caption{\label{meaneds}
The present mean value, $\overline{\left\langle\delta\theta\right\rangle}/3H$, 
in the matter-dominated Universe as a function of the ultraviolet cutoff 
$k_\textrm{MAX}$ for various values of the shape factor $\Gamma$.}
\end{figure}

The result for the present mean value of
$\overline{\left\langle\delta\theta\right\rangle}/3H$ is shown in
Fig.\ \ref{meaneds} for various choices of $\Gamma$.  It scales as
$a/a_0=1/(1+z)$.

\begin{figure}
\includegraphics[width=0.9\textwidth]{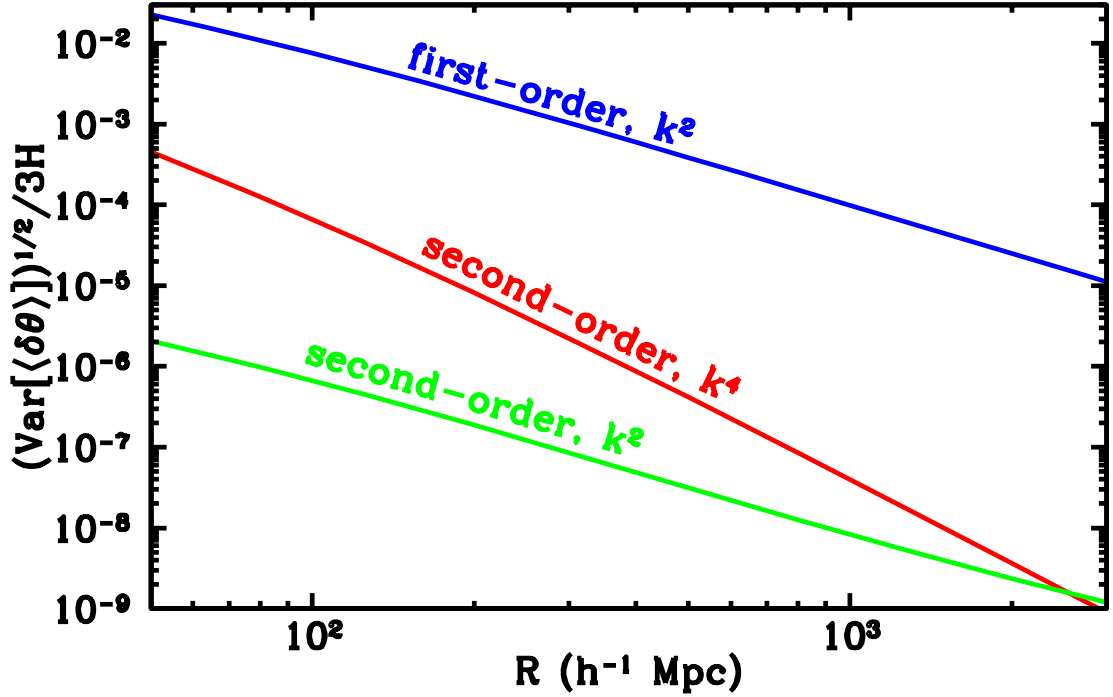}
\caption{\label{variance}
The present contribution to $\sqrt{\textrm{Var}\left[
\langle\delta\theta\rangle\right]}/3H$ from the indicated terms
in a matter-dominated Universe as a function of $R$, the volume
sampled.  The results are for $k_\textrm{MAX}=0.1$ and $\Gamma=0.2$.
On the logarithmic scale, the dependence on $\Gamma$ and
$k_\textrm{max}$ is not noticeable. We emphasize the point made in the
text that the calculation for the the second-order, two-derivative term
has an infrared cutoff of the present Hubble radius.}
\end{figure}

Now consider the variance about the mean value.  
First consider the variance of the linear term, Eq.\
(\ref{varlinear}).  This term contributes to the variance in
$\left\langle\delta\theta\right\rangle/3H$ an amount
\begin{equation}
\left.\frac{\sqrt{\textrm{Var}\left[\langle\delta\theta\rangle\right]}}{3H}
\right|_\textrm{linear term}=
\frac{5}{9}\frac{1}{a^2H^2}\sqrt{\textrm{Var}
\left[\left\langle\nabla^2\varphi\right\rangle_1\right]} =
\frac{5}{6} \left[\int_0^\infty \frac{dk}{k} \Delta^2(k) \ W^2(kR) 
\right]^{1/2}.
\end{equation}
Defining a dimensionless wavenumber $x=q\Gamma r$, the term becomes
\begin{equation}
\left.\frac{\sqrt{\textrm{Var}\left[\langle\delta\theta\rangle\right]}}{3H}
\right|_\textrm{linear term} =
\frac{5}{6}A \left(\frac{h\, \textrm{Mpc}^{-1}}{H_0}\right)^2\frac{a}{a_0} 
 r^{-2}\left[ \int_0^{x_\textrm{MAX}} dx \ x^3 \ T^2(x/\Gamma r) \ 
e^{-x^2} \right]^{1/2} \simeq \frac{a}{a_0}\frac{100}{r^2}\quad
(\Gamma r\gg1),
\end{equation}
where the last expression holds for $x_\textrm{MAX}=k_\textrm{MAX}R\gg
1$. The result is shown in Fig.\ \ref{variance}.  Because of the
window function, the results do not depend on $k_\textrm{MAX}$, the
ultraviolet cutoff (so long as it is greater than about $k=0.1
h\textrm{ Mpc}^{-1}$).  The result is also well behaved in the
infrared. Of course as $R\rightarrow\infty$, the variance of the linear term
approaches the mean of the linear term, which is zero.

We can see comparing Fig.\ \ref{meaneds} and Fig.\ \ref{variance} that
the variance in the linear term dominates the mean value out to
distances of about $3\times10^3h^{-1}$ Mpc.

Next, consider the contribution to the variance of a typical second-order,
four-derivative term, Eq.\ (\ref{phi2k4}).   This term contributes to
the variance in $\left\langle\delta\theta\right\rangle/3H$ an amount
\begin{equation}
\left.\frac{\sqrt{\textrm{Var}\left[\langle\delta\theta\rangle\right]}}{3H}
\right|_{\textrm{second-order, }k^4}=
\frac{4}{27}\frac{1}{a^4H^4}\sqrt{
\textrm{Var} \left[ \left\langle\left(\nabla^2\varphi\right)^2\right\rangle
- \left\langle\varphi_{,ij}\varphi^{,ij}\right\rangle\right]}.
\end{equation}
Again, with the same dimensionless variables 
\begin{eqnarray}
\left.\frac{\sqrt{\textrm{Var}\left[\langle\delta\theta\rangle\right]}}{3H}
\right|_{\textrm{second-order, }k^4} & = & \frac{1}{3}A^2
\left(\frac{h\,\textrm{Mpc}^{-1}}{H_0}\right)^4\left(\frac{a}{a_0}\right)^2 
r^{-4} \left[ \int_0^{x_\textrm{MAX}} dx_1\int_0^{x_\textrm{MAX}} dx_2 \ x_1^3 
\ x_2^3  \right. \nonumber \\
& & \left. \times \ T^2(x_1/\Gamma r) \ T^2(x_2/\Gamma r) \
e^{-x_1^2}e^{-x_2^2}\int_{-1}^1 d\mu \ e^{-2x_1x_2\mu}\ 
\left(1-\mu^2\right)^2 \right]^{1/2} .
\end{eqnarray}
The results are also shown in Fig.\ \ref{variance}.  This term scales
as $r^{-4}$. The result is well behaved in the infrared and an
infrared cutoff does not need to be introduced.

Finally, consider the variance of a typical second-order two-derivative
term, Eq.\ (\ref{phi2k2}).  As mentioned in the previous section this term is
not well behaved in the infrared.  We first calculate the variance by
introducing an infrared cutoff which we will take to be the wavenumber of the
present Hubble radius.   The second-order $k^2$ term contributes to the 
variance in $\left\langle\delta\theta\right\rangle/3H$ an amount
\begin{equation}
\left.\frac{\sqrt{\textrm{Var}\left[\langle\delta\theta\rangle\right]}}{3H}
\right|_{\textrm{second-order, }k^2}=
\frac{25}{27}\frac{1}{a^2H^2}
\sqrt{\textrm{Var}\left[ \left\langle\varphi\nabla^2\varphi\right\rangle +
\frac{13}{18}\left\langle\varphi_{,i}\varphi^{,i}\right\rangle\right] } .
\end{equation}
In terms of the dimensionless variables $x$ and $r$, this term contributes to
the variance an amount
\begin{eqnarray}
\left.\frac{\sqrt{\textrm{Var}\left[\langle\delta\theta\rangle\right]}}{3H}
\right|_{\textrm{second-order, }k^2} & = & \frac{25}{24}
A^2 \left(\frac{h\,\textrm{Mpc}^{-1}}{H_0}\right)^2\frac{a}{a_0} 
r^{-2} \left[ \int_{x_\textrm{MIN}}^{x_\textrm{MAX}} dx_1
\int_{x_\textrm{MIN}}^{x_\textrm{MAX}} dx_2  
\ T^2(x_1/\Gamma r) \ T^2(x_2/\Gamma r) \ e^{-x_1^2}e^{-x_2^2}\right. 
\nonumber \\
& &  \times \left. 
\int_{-1}^1 d\mu \ e^{-2x_1x_2\mu} \ \left(\frac{x_1^3}{x_2}+\frac{x_2^3}{x_1}
+  2x_1 x_2 + \frac{26}{9}x_1^2\mu + \frac{26}{9}x_2^2\mu 
+\frac{169}{81}x_1x_2\mu^2\right)  \right]^{1/2} .
\end{eqnarray}
As a result of the infrared behavior, for this term we must introduce
another parameter, $x_\textrm{MIN}=k_\textrm{MIN} R$, as an infrared cutoff. 
Its value is shown in Fig.\ \ref{variance} as a function of $R$.  As expected,
it has a $r^{-2}$ behavior. 

Now we turn to the infrared-singular part of the second-order
two-derivative term. The contribution is
\begin{equation}
\left.\frac{\sqrt{\textrm{Var}\left[\langle\delta\theta\rangle\right]}}{3H}
\right|_{\textrm{second-order, }k^2}=
\frac{25}{27}\frac{1}{a^2H^2}
\sqrt{\left[ \left\langle\varphi\nabla^2\varphi\right\rangle 
-3 \left\langle\varphi\right\rangle_1 
\left\langle\nabla^2\varphi\right\rangle_1\right]_\infty} .
\end{equation}
We will evaluate the integral by defining some $x_*=k_*R$
such that $x_\textrm{MIN} \ll x_* < 1$. We will be interested in large
values of $r$, so if $x\ll 1$, then $x/\Gamma r \ll 1$, and
$T^2(x/\Gamma r)$ can be set to unity.
Using Eq.\ (\ref{blowup}), this term may be written as
\begin{equation}
\left.\frac{\sqrt{\textrm{Var}\left[\langle\delta\theta\rangle\right]}}{3H}
\right|_{\textrm{second-order, }k^2}=\frac{25}{12} A^2
\left(\frac{h\,\textrm{Mpc}^{-1}}{H_0}\right)^2\frac{a}{a_0} 
 r^{-2}\left[ x_*^4\ln \frac{k_*}{k_\textrm{MIN}}\right]^{1/2} .
\end{equation}
If we take $k_*= k_H$ where $k_H=H=3000^{-1}h 
\textrm{ Mpc}^{-1}$ is the Hubble wavenumber, then evaluating the term for $r$
corresponding to the Hubble radius we find
\begin{equation}
\left.\frac{\sqrt{\textrm{Var}\left[\langle\delta\theta\rangle\right]}}{3H}
\right|_{\textrm{second-order, }k^2}  
= \frac{25}{24}A^2\frac{a}{a_0}  
\left[\ln\frac{k_H}{k_\textrm{MIN}}\right]^{1/2} 
= \frac{25}{36}A\frac{a}{a_0}\sqrt{\textrm{Var}[\langle\varphi\rangle_1]}.
\label{logterm}
\end{equation}
We started with a perturbative expansion, and for that expansion 
to be valid requires $\varphi<1$, so our perturbative analysis will break down
for $\sqrt{\textrm{Var}[\langle\delta\theta\rangle]}/3H \sim 10^{-5}$.

Of course the total variance of $\langle\delta\theta\rangle/3H$
includes cross terms from the various contributions.  However we have
seen that the linear term will dominate.

\section{Conclusions}

Our results are illustrated in Figs \ref{meaneds} and \ref{variance}.  The mean
corrections to the expansion rate are a few parts in $10^5$. If the correct
prescription for dealing with the infrared singular nature of
$\textrm{Var}[\varphi]$ is to employ a cutoff of the order of the Hubble
radius, then the variance about the mean is small, and dominated by the
first-order term. If, however, the super-Hubble modes are physical, then the
variance is dominated by the infrared singular part of the second-order
corrections. Restricting ourselves to $\textrm{Var}[\varphi]<1$ where the
perturbative calculation is valid, then the variance will be small, of order
$10^{-5}$.

In Ref.\ \cite{Rasanen}, R\"as\"anen considered what amounts to the variance
associated with the second-order term proportional to
$\nabla^2\varphi\nabla^2\varphi$ and suggested that due to unknown boundary 
conditions it may lead to a large contribution to the
expansion rate.  We claim that when properly averaged, this term is
subdominant.

In closing, let us speculate that corrections to the expansion rate will
include a term like $\varphi\nabla^2\varphi$ even when $\varphi\gg 1$.
Returning to the infrared-singular part of the second-order variance, Eq.\
(\ref{logterm}),
\begin{equation}
\left.\frac{\sqrt{\textrm{Var}\left[\langle\delta\theta\rangle\right]_\infty}}
{3H} \right|_{\textrm{second-order, }k^2}  =
3.76\times10^{-10} \frac{a}{a_0}\ln\frac{k_H}{k_\textrm{MIN}} .
\label{logterm2}
\end{equation}
For the variance to be of order unity, the perturbation spectrum
would have to extend to a factor of $\exp(6\times10^{18})$ ($10^{18.8}$
$e$-folds!) times the present Hubble radius.   

Rather than a Harrison--Zel'dovich spectrum, if we assume a slightly
red spectrum so that $\Delta^2(k)\propto k^{4-\epsilon}$ with $0 <
\epsilon \ll 1$,\footnote{In the the usual parlance, this corresponds
to a scalar spectral index of $1-n = \epsilon$.} then the logarithmic term in
Eq.\ (\ref{logterm2}) is replaced by $\epsilon^{-1}
(k_H/k_\textrm{min})^\epsilon$. Now this will give unit variance if $\ln
k_H/k_\textrm{MIN}=(43.3+\ln \epsilon)/\epsilon$.  For instance, if $n=0.94$ on
super-Hubble-radius scales, then a variance of order unity is obtained if the
perturbation spectrum extends 676 $e$-folds beyond the Hubble radius. Since the
present Hubble radius corresponds to a scale that crossed the Hubble radius
about 60 $e$-folds before the end of inflation, if inflation lasted more than
736 $e$-folds with a super-Hubble-radius spectral index of $n=0.94$, then the
effect of super-Hubble-radius perturbations would have a large effect on the
expansion rate of our Hubble volume.  Speculation about the significance of
this result will follow in a separate paper \cite{kmnrtwo}.

\section*{Acknowledgments}

E.W.K.\ is supported in part by NASA grant NAG5-10842 and by the Department of
Energy.  We would like to thank Enrico Barausse, Riccardo Catena, 
Scott Dodelson, Lam Hui, Luigi Pilo, Syksy R\"as\"anen, and Dominik Schwarz 
for helpful conversations.

\appendix

\section{On the prescription for $T^{(2)}_{\mu\nu}$ and
$g^{(2)}_{\mu\nu}$ \label{alternative}}

The aim of this appendix is to demonstrate that $\langle\theta\rangle$
does not depend on the value of $\langle g^{(2)}_{\mu\nu} \rangle$
since second-order terms will only enter as spatial derivatives.  For
the demonstration, we first introduce $\vartheta^i_{\ j}$, the
extrinsic curvature of constant-$\tau$ hypersurfaces:
\begin{equation}
\vartheta^i_{\ j} = \frac{1}{2}\gamma^{ik}\gamma^\prime_{kj}.
\end{equation}
In either the synchronous or comoving gauge one can see that 
(see Ref.\ \cite{MMB}) 
\begin{equation}
\theta^{(1)}+\theta^{(11)}+\theta^{(2)}=\frac{\vartheta^i_{\ i}}{a} \equiv
\frac{\vartheta} {a}  .
\label{vartheta}
\end{equation}
One can now write the $G^0_{\ 0}$ term in a convenient form (see Eq.\
(4.3) of \cite{MMB})
\begin{equation}
 G_{00}^{(1)}+ G_{00}^{(11)}+ G_{00}^{(2)}
= \frac{1}{2} \left(4 \frac{a'}{a} \vartheta + ^3\!{\cal R} 
  + \vartheta^2-\vartheta^i_{\ j}  \vartheta^j_{\ i} \right)  , 
\label{eqR}
\end{equation}
where $^3{\cal R}$ is the intrinsic curvature of the three space
with metric $\gamma_{ij}$. At second order $^3{\cal R}$ is
\begin{eqnarray}
^3{\cal R}  & = & \gamma^{(1)\ell k}_{~~~~~,\ell k} -
\gamma^{(1)k~,\ell}_{~k~~,\ell}
+ \frac{1}{2} \left( \gamma^{(2)\ell k}_{~~~~~,\ell k} -
\gamma^{(2)k~,\ell}_{~k~~,\ell} \right) + \gamma^{(1)jk}
\left( \nabla^2 \gamma^{(1)}_{jk}
+ \gamma^{(1)\ell}_{~\ell~,jk} - 2 \gamma^{(1)\ell}_{~j~,\ell k} \right)
\nonumber\\
& &
+ \gamma^{(1)\ell k}_{~~~~,\ell}
\left( \gamma^{(1)j}_{~j~,k} - \gamma^{(1),j}_{jk} \right)
+ \frac{3}{4} \gamma^{(1)\ell j}_{~~~~,k} \gamma^{(1),k}_{~\ell j}
- \frac{1}{2} \gamma^{(1)\ell j}_{~~~~,k} \gamma^{(1)k}_{~j~,\ell}
- \frac{1}{4} \gamma^{(1)j,\ell}_{~j} \gamma^{(1)k}_{~k~,\ell} .
\label{R}
\end{eqnarray}
Using Eq.\ (\ref{vartheta}), Eq.\ (\ref{eqR}) also can be written  as
 \begin{equation}
G_{00}^{(1)}+G_{00}^{(11)}+G_{00}^{(2)} = \frac{1}{2} \left[4a'
\left(\theta^{(1)}+\theta^{(11)}+\theta^{(2)}\right)
+^3{\cal R}^{(1)}+^3\!{\cal R}^{(11)}+^3\!{\cal R}^{(2)} +
  (\vartheta^{(1)})^{2}-\vartheta^{(1)i}_{\ \ \ \ j}
    \vartheta^{(1)j}_{\ \ \ \ i}  \right] .
\label{usefulrel}
\end{equation}

Recall now that in any of the definitions adopted, Eq.\ (\ref{combo3})
is always valid.  So independent of $\langle g^{(2)}_{\mu\nu} \rangle$,
\begin{equation}
\langle \theta \rangle = 3 H - \frac{\left\langle G_{00}^{(1)}+
G_{00}^{(11)}+G_{00}^{(2)} \right\rangle}{2 a^2H} +
\left\langle \theta^{(1)}\right\rangle +\left\langle
\theta^{(11)}\right\rangle
+\left\langle \theta^{(2)} \right\rangle .
\end{equation}
Using Eq.\ ({\ref{usefulrel}}), we obtain
\begin{equation}
\langle \theta \rangle = 3 H - \frac{1}{4 a^2 H} \left[
\left\langle ^3{\cal R}^{(1)} + ^3\!{\cal R}^{(11)} \right\rangle
+ \left\langle ^3{\cal R}^{(2)} \right\rangle
+  \left\langle (\vartheta^{(1)})^{2}-\vartheta^{(1)i}_{\ \ \ \ j}
    \vartheta^{(1)j}_{\ \ \ \ i}
  \right\rangle \right]  .
\label{complete}
\end{equation}
We note now that the only piece that depends on $g_{\mu \nu}^{(2)}$
is the one contained in $^3{\cal R}^{(2)}$. And, as we can see from its
explicit expression
\begin{equation}
^3{\cal R}^{(2)} = \frac{1}{2}\left(\gamma^{(2)lk}_{\ \ \ \ \ ,lk}
                -\gamma^{(2)kk}_{\ \ \ \ \ ,ll} \right) , 
\end{equation}
this is a spatial gradient, so independent on  $\langle g_{\mu
  \nu}^{(2)}\rangle$.
From this formula it is also clear that $\theta$ has no term with zero
powers of $k$, since the spatial curvature $^3{\cal R}$ always
involves  spatial derivatives (and the same is true for the
$\left\langle(\vartheta^{(1)})^2\right\rangle
- \left\langle \vartheta^{(1)i}_{\ \ \ \ j}\vartheta^{(1)j}_{\ \ \ \ i}
\right\rangle$ terms).

If one computes $\langle\theta\rangle$ using Eq.\ (\ref{complete}),
one obtains the same result as Eq.\ (\ref{risultatoWithBt}), thus our
results are independent of $\langle g^{(2)}_{\mu\nu} \rangle$.  By the same
argument, in a first-order calculation we could add a constant to 
$g^{(1)}_{\mu\nu}$.

\section{$\theta$ in the Poisson gauge \label{thetapoisson}}

In this appendix we repeat the calculation in the Poisson gauge in
order to check our results and compare with other results in the
literature for $\theta$ in the that gauge.\footnote{Actually, the
other results we are aware of in the literature use only the
first-order metric.}  The Poisson gauge is a second-order
generalization of the longitudinal (or Newtonian) gauge.  Since this
is a very involved way to do the computation, and since it is only
intended to be a check, we do it only in a partial way by just
computing the corrections $\theta^{(1)}+\theta^{(11)}+\theta^{(2)}$
[see Eqs.\ (\ref{combo11}) and (\ref{fundamentalsyn})]. 
Furthermore we compute exactly the terms with two spatial
derivatives, while as for the four derivatives terms, we only check
that they have zero statistical mean (that is, we disregard total
spatial gradients).

The metric is given at second order in this gauge by\footnote{Also
here, for our purposes the vector $\omega_i^{(2)}$ and the tensor
$\chi_{ij}^{(2)}$ will never enter in the computations.}
\begin{equation}
 ds^2 = a_P^2(\eta)\left\{-\left(1+2\phi^{(1)}+\phi^{(2)}\right)d\eta^2 +
\omega^{(2)}_i dy^i d\eta+ \left[
\left(1-2\psi^{(1)}-\psi^{(2)}\right) \delta_{i j}+\frac{1}{2}
 \chi_{i j}^{(2)} \right]  dy^i dy^j \right\} .
\end{equation}
Here $\chi_{i j}^{(2)}$ is a pure symmetric, transverse, traceless
tensor degree of freedom.  We will denote the derivative
with respect to conformal time $\eta$ as $\partial_{\eta}$.  We will
also introduce here a cosmic time $z$, linked to $\eta$ through the
relation $ dz= a \, d\eta $, and we will denote the derivative with
respect to $z$ with a $\partial_z$ symbol.  The spatial derivatives
with respect to $x^i$ will be denoted as before, with a simple
$\partial^i$ or $^{, i}$ symbol. The derivatives with respect to $y^i$
instead will be explicitly written as $\partial/\partial y^i$ or with
the subscript $\partial_{{\bf (y)}}^i$.  Recall also, as well known,
that the $i-i$ component of Einstein's equations (in the absence of
anisotropic stress) imposes $\phi^{(1)}=\psi^{(1)}$, and moreover, the
evolution equations in matter domination give
\begin{equation}
\phi^{(1)}(z,y^i)=\psi^{(1)}(z,y^i)=\varphi(y^i) .
\end{equation}
The four velocity of the fluid $u^{\mu}$ here has a more involved form
\begin{subequations}
\begin{eqnarray}
u^0&=&\frac{1}{a_P}\left( 1-\varphi + \frac{3}{2}\varphi^2+
\frac{2}{9 {\cal H}_P^2}
\partial_{{\bf (y)}\, i}\varphi \partial_{{\bf (y)}}^i \varphi
-\frac{\phi^{(2)}}{2} \right)  ,\\
u^{i}&=&\frac{1}{a_P}\left (-\frac{2}{3 {\cal H}_P} \partial_{{\bf (y)}}^i
\varphi + \frac{v^{(2) \, i}}{2} \right) ,  \label{ui1}
\end{eqnarray}
\end{subequations}
where ${\cal H}_P\equiv a_P^{-1}\partial_{\eta}a_P$.  Here $v^{(2)\,
i}$ is the second-order contribution to the three velocity.  We will
need only its divergence (which is obtained taking the divergence of
the $0-i$ Einstein equation):
\begin{eqnarray}
\partial_{{\bf (y)} \, i} v^{(2)\,  i} &=&
\frac{4}{3 {\cal H}_P} \left(
-\partial_{{\bf (y)}}^i \varphi \partial_{{\bf (y)}_i} \varphi
-\varphi \nabla^2_{{\bf (y)}} \varphi
+\frac{2(\partial_{{\bf (y)}}^i \varphi \nabla^2_{{\bf (y)}}
\partial_{{\bf (y)} \, i}  \varphi)}{3 {\cal H}_P^2}
+ \frac{2( \nabla^2_{{\bf (y)}} \varphi)^2 }{3 {\cal H}_P^2}  \right)
\nonumber \\
&-&\frac{2}{3 {\cal H}_P} \left( \nabla_{{\bf (y)}}^2 \phi^{(2)}
+\frac{\nabla_{{\bf (y)}}^2 \psi^{(2)\prime}}{\cal{H}} \right) \, .
\label{ui2}
\end{eqnarray}
Now, taking these expressions and taking Christoffel symbols from Ref.\
\cite{Acquavivaetal}, we must compute
\begin{equation}
\label{thetapoiss}
\theta=\partial_{\mu} u^{\mu}+\Gamma_{\mu \alpha}^{\alpha} u^{\mu}=
 \partial_{\eta}u^{0}+ \partial_{{\bf (y)} \, i} u^i+\Gamma^\alpha_{\mu
\alpha}   u^{\mu}  .
\end{equation}
This gives us
\begin{equation}
\theta(z,y^i) = 3 H_P-3 H_P \varphi + \frac{9 H \varphi^2}{2}+
\frac{20 \partial_{{\bf (y)} \, i} \varphi \partial_{{\bf (y)}}^i \varphi}
{9 a_P^2 H_P} -\frac{3 H \phi^{(2)}}{2} -\frac{3 \dot{\psi}^{(2)}  }{2} +
\partial_{{\bf (y)} \, i} u^{(1) \, i}+\partial_{{\bf (y)} \, i} u^{(2)\,i} .
 \label{thetaexpl}
\end{equation}
Here  $H_P$ is defined as $H_P\equiv a_P^{-1}\partial_z a_P=
a_P^{-2}\partial_{\eta}a_P=a_P^{-1} {\cal H}_P$.
The piece coming from $\partial_{{\bf (y)} \, i} u^i$ in Eq.\
(\ref{thetapoiss})
contains first order ($\partial_{{\bf (y)} \, i} u^{(1) \, i}$) and
second  order  ($\partial_{{\bf (y)} \, i} u^{(2) \, i}$) terms.
The first order term [see Eq.\ (\ref{ui1})] is equal to
\begin{equation}
\partial_{{\bf (y)} \, i} u^{(1) \, i} =
-\frac{2}{3 a_P^2 H_P } \nabla^2_{{\bf (y)}} \varphi   ,   \label{u1}
\end{equation}
and it will be important in the next steps of the computation, since
it will produce second-order quantities.  The term $\partial_{{\bf
(y)} \, i} u^{(2) \, i}$ [Eq.\ (\ref{ui2})] is a spatial gradient of
second-order quantities. For the purposes of this appendix we have to
keep the first two pieces in Eq.\ (\ref{ui2}), and moreover we have to
consider also the two intrinsically second-order terms $\phi^{(2)}$
and $\psi^{(2)}$. Note also that, in their time evolution they contain
non-local quantities ($\Theta_0$ and $\Psi_0$ of Eqs.\ (6.8) of Ref.\
\cite{MMB}), {\it i.e.,} they are defined through
\begin{eqnarray}
\nabla^2_{{\bf (y)}} \Theta_0 &=& \Psi_0-\frac{1}{3} \partial_{{\bf (y)}
\, i} \varphi \partial_{{\bf (y)}}^i \varphi  , \nonumber \\
\nabla^2_{{\bf (y)}} \Psi_0 &=& -\frac{1}{2}
\left[ ( \nabla^2_{{\bf (y)}} \varphi)^2-
\partial_{{\bf (y)} \, j} \partial_{{\bf (y)} \, k} \varphi
\partial_{{\bf (y)}}^j \partial_{{\bf (y)}}^k \varphi \right]  .
\label{Theta_0}
\end{eqnarray}
So the spatial derivatives of these quantities will produce not only
spatial gradients, but also terms who do not have zero statistical
mean.  In the end, we keep them in the calculation as
\begin{equation}
\partial_i u^{(2)\,i}=-\frac{\nabla^2_{{\bf (y)}} \phi^{(2)}}{3 a_P^2 H_P}-
\frac{\nabla^2_{{\bf (y)}} \dot{\psi}^{(2)}}{3 a_P^2 H_P^2}
-\frac{2}{3 a_P^2 H_P} \left(
\partial_{{\bf (y)}}^i \varphi \partial_{{\bf (y)}_i} \varphi
+\varphi \nabla^2_{{\bf (y)}} \varphi \right) + \cdots   , \label{u2}
\end{equation}
where ``$\cdots$'' indicates terms with four spatial gradients 
(that we will omit).

So, the relevant expression for $\theta(z,y^i)$ is obtained by Eqs.\
(\ref{thetaexpl}), (\ref{u1}), and (\ref{u2})
\begin{eqnarray}
\theta(z,y^i) &=& 3H_P-3 H_P \varphi + \frac{9 H \varphi^2}{2} +
\frac{20 \partial_{{\bf (y)} \, i} \varphi \partial_{{\bf (y)}}^i \varphi}
{9 a_P^2 H_P} -\frac{2}{3 a_P^2 H_P} \left(
\partial_{{\bf (y)}}^i \varphi \partial_{{\bf (y)}_i} \varphi
+\varphi \nabla^2_{{\bf (y)}} \varphi \right) \nonumber \\
&-&\frac{3 H \phi^{(2)}}{2}
- \frac{3 \dot{\psi}^{(2)} }{2} 
- \frac{2 \nabla^2_{{\bf (y)}} \varphi}{3 a_P^2 H_P}
- \frac{\nabla^2_{{\bf (y)}} \phi^{(2)}}{3 a_P^2 H_P}
- \frac{\nabla^2_{{\bf (y)}}\dot{\psi}^{(2)}}{3 a_P^2 H_P^2} + \cdots .
\label{thetatx}
\end{eqnarray}
This is the result in the Poisson gauge in coordinates $(z,y^i)$. Note
that here there are terms which do not vanish in the infrared limit,
proportional to $\varphi$ and in $\varphi^2$ (they will disappear
going back to the synchronous and comoving coordinates, as noted in
Ref.\ \cite{geshzjiani}).

In order to compare with the results already obtained in the
synchronous gauge, we have to express our quantity $\theta(z,y^i)$ as
$\theta(t,x^i)$ and average it over a volume in coordinates $x^i$ at
constant $\tau$ as in Section \ref{thetasynchronous}.  First, we have
to change time from $z$ to $t$ in all the quantities.  This is
relevant only for the zeroth-order and first-order terms, and not for
the ones which are already second order.  So the quantities in which
the time has to be changed are $H_P$, $a_P(z)$ and $\varphi(z,y^i)$.
The first two, $ a_P(z)=\left(z/z_0\right)^{2/3}$ and $H_P=2/3 z$,
have to be expanded up to second order in $\delta t$, which is defined
by $t=z+\delta t \, $ and expressed in terms of the scale factor and
the Hubble rate defined by a comoving observer: $a=\left( t/z_0
\right)^{2/3}$, $H=2/3 t$.  Instead, for the variable
$\varphi(\eta,y^i)$, $\varphi(\eta,y^i)=\varphi(\tau,x^i)$ holds since
$\partial_{\eta}\varphi(\eta,y^i)=0$.  In order to find the change of
the time coordinate from $z$ to $t$ one knows that $u^{\mu}
\partial_{\mu}=\partial_{t}$, and that $\partial_{t} t=1$. In this way
one obtains an iterative equation
\begin{equation}
\left( 1+v^{(1) \, 0}+v^{(2) \, 0}\right) \frac{d}{dz} \left(z+\delta
t^{(1)} + \delta t^{(2)}\right)
+ u^{(1) \, i} \partial_{{\bf (y)} \, i} \left(z+\delta t^{(1)}\right)=1 ,
\end{equation}
where $u^{\mu}\equiv a_P^{-1}(\delta^{\mu}_0 +v^{(1)}+v^{(2)})$.  For
zeroth, first, and second order, the equation becomes
\begin{eqnarray}
1 & = & \frac{dz}{dz}  \nonumber \\
0 & = & v^{(1) \, 0}+ \frac{d}{dz}\delta t^{(1)}  \nonumber \\
0 & = & v^{(2) \, 0} + v^{(1) \, 0} \frac{d}{dz}\delta t^{(1)} +
\frac{d}{dz}\delta t^{(2)}+
u^{(1) \, i} \partial_{{\bf (y)} \, i} \delta t^{(1)}  ,
\end{eqnarray}
which lead to
\begin{eqnarray}
\delta t^{(1)}& = & \varphi t  \label{deltatau} \nonumber \\
\delta t^{(2)}& = & -\frac{1}{2}\varphi^2 t+\frac{4 \varphi_{, i }
\varphi^{, i}}{45 a^2 H } + \frac{1}{2}\int^{t} d t' \phi^{(2)}  .
\end{eqnarray}

Now using these equations, one can change the time in Eq.\
(\ref{thetatx}), obtaining
\begin{equation}
\theta(t,y^i)=3 H+
\frac{118 \varphi^{{\bf (y)},i} \varphi_{{\bf (y)},i} }{45 a^2 H}
-\frac{3 H \phi^{(2)}}{2} -\frac{3 \dot{\psi}^{(2)}}{2}
+\frac{9 H^2 \int^{\tau} \phi^{(2)} }{4} + \partial_{{\bf (y)} \, i} u^{i} .
\label{thetataux}
\end{equation}
As said before, this transformation ({\it i.e.,} passing to
synchronous time $t$) has eliminated the two terms in $\varphi$ and
$\varphi^2$ that do not vanish in the infrared.  We wrote it in this
form in order to compare with a previous result in literature (Eq.\
(10) of Ref.\ \cite{Rasanen}). As we are going to show, we agree with
that result, but we added here the contribution of intrinsic
\footnote{Note however that what is ``intrinsic'' second order in a
gauge, can become first order times first order in another one.} 
second-order terms $\phi^{(2)}$ and $\psi^{(2)}$.  This is crucial if
we want to recover the result obtained in Sec.\ \ref{thetasynchronous}
in the other gauge.  Here $\partial_{{\bf (y)} \, i} u^{i}$ contains
the terms
\begin{equation}
\partial_{{\bf (y)} \, i} u^{i}=-\frac{2 \nabla^2_{{\bf (y)}} \varphi  }
{3 a^2 H} - \frac{2 \varphi \nabla^2_{{\bf (y)}} \varphi  }{9 a^2 H} 
- \frac{2}{3 a^2H} \left(
\partial_{{\bf (y)}}^i \varphi \partial_{{\bf (y)}_i} \varphi
+\varphi \nabla^2_{{\bf (y)}} \varphi \right)
 - \frac{\nabla^2_{{\bf (y)}} \phi^{(2)}}{3 a^2 H}
 - \frac{\nabla^2_{{\bf (y)}} \dot{\psi}^{(2)}   }{3 a^2 H^2} + \cdots .
\end{equation}
Note that after changing time, a new term (which is not a spatial
gradient) appears, namely $-2 \varphi \nabla^2_{{\bf (y)}} \varphi/9
a^2 H$.

Finally, one has to change also spatial variables from $y^i$ to $x^i$.
The change of coordinates is found by imposing that the metric
$g_{\mu\nu}$ in the coordinates $(t,x^i)$ has $g_{0 i}=0$.  Using the
transformation given by Eq.\ (\ref{deltatau}), one finds that to first
order the change in spatial coordinates is given by
\begin{equation}
x^i=y^i+ \frac{2}{3} \frac{\partial_{{\bf (y)}}^i \varphi}{a^2 H} .
\label{transformation}
\end{equation}
Since it is already first order, this change of coordinates is
irrelevant for quantities that are already second-order.  Moreover,
the only first order term of Eq.\ (\ref{thetataux}) is $
\partial_{{\bf (y)} \, i} u^{(1) \, i}= -2 \nabla^2_{{\bf (y)}}
\varphi/3 a^2 H$.  Here, the change of spatial coordinates in the
argument of $\varphi(\tau,y^i)$ is not relevant for our purposes,
since it produces only an extra term which is a four derivative
spatial gradient of something.  But passing from $\nabla^2_{{\bf (y)}}
$ to $\nabla^2$ produces a (relevant) extra term with four spatial
derivatives,
\begin{equation}
- \frac{2}{3  a^2 H} \nabla^2_{{\bf (y)}} \varphi=
- \frac{2}{3 a^2 H } \nabla^2 \varphi -
  \frac{4\,{\varphi }^{,j\,k}\,{{\varphi }_{,j\,k}}} {9\, a^4\, H^3}   .
\end{equation}
In all the other terms we can safely put $\partial_{{\bf (y)}}=
\partial$.  After doing this, the result is
\begin{eqnarray}
\theta(t,x^i)&=& 3H - \frac{2 \nabla^2 \varphi}{3 a^2 H}
-\frac{8 \varphi \nabla^2 \varphi}{9 a^2 H} +
\frac{88 \varphi^{,i} \varphi_{,i} }{45 a^2 H}
 - \frac{4\,{\varphi }^{,j\,k}\,{{\varphi }_{,j\,k}}}
   {9\,\,a^4\,{H}^3} - \nonumber \\
&&
 \frac{3 H {\phi}^{(2)}  }{2} - \frac{3 \dot{\psi}^{(2)}  }{2}
+ \frac{9 H^2 \int^{t} {\phi}^{(2)}  }{4}
-\frac{\nabla^2 \phi^{(2)}}{3 a^2 H}-
\frac{\nabla^2 \dot{\psi}^{(2)}}{3 a^2 H^2} + \cdots{} .
 \label{thetatauy}
\end{eqnarray}
Using Eqs.\ (6.8) of Ref.\ \cite{MMB}, together with Eqs.\
(\ref{Theta_0}) given in the present section, one obtains for the time
evolution of the intrinsic second order terms $\phi^{(2)}$ and
$\psi^{(2)} $ the result
\begin{equation}
\theta(t,x^i)= 3H - \frac{2 \nabla^2 \varphi}{3 a^2 H}
+\frac{5 \varphi^{,i} \varphi_{,i} }{3 a^2 H}
-\frac{40 \varphi \nabla^2 \varphi}{9 a^2 H}
-\frac{4\,{\varphi }^{,j\,k}\,{{\varphi }_{,j\,k}}}
   {9\,\,a^4\,H^3} + \cdots .
\label{thetaPf}
\end{equation}
Now this is the first check we wanted to do.  We can compare with Eq.\
(\ref{fundamentalsyn}) and see that deriving by parts and using $a
H=2/\tau$, they coincide (up to four-derivative total spatial
gradients).

An important observation here is that the presence of the
intrinsically second-order terms is necessary to obtain not only the
correct result, but even a consistent calculation.  In fact if one
ignores them, {\it i.e.,} ignores $\psi^{(2)}$ and $\chi^{(2)}$ in the
synchronous gauge and ignores $\phi^{(2)}$ and $\psi^{(2)}$ in the
Poisson gauge, and compares the two results, one sees that they do not
coincide.  In other words they are not really ``intrinsic'' second
order, since what appears as intrinsic second order in one gauge, can
appear as first order times first order in another gauge.

At this point we may also average expression Eq.\ (\ref{thetaPf}) over
a volume $d^3\!x$ as we did in Sec.\ \ref{thetasynchronous}.  This has
to be done with the appropriate measure of integration, which is
$d^3\!x \sqrt{\gamma}$, where $\gamma \equiv \textrm{det}(g_{ij})$.
Up to first order the spatial metric is given by
\begin{equation}
g_{ij}=\frac{\partial y^{\alpha}}{\partial x^i}\frac{\partial y^{\beta}}
{\partial x^j}  \tilde{g}_{\alpha \beta}
=a_P^2 \left[\delta_{ij}(1-2\varphi) -\frac{4}{3 a^2
H^2}{\partial_i\partial_j \varphi} \right]
=a^2\left(1-\frac{4}{3}\varphi\right)  \left[\delta_{ij}(1-2\varphi)
 -\frac{4}{3 a^2 H^2}{\partial_i\partial_j \varphi} \right] \, ,
\end{equation}
where $\tilde{g}$ is the metric in the $z,y$ coordinates, and where we
used the transformation of Eq.\ (\ref{transformation}).  Using this
metric, the measure becomes
\begin{equation}
d^3\!y \sqrt{\gamma}= d^3y \, a^3 \left[1-5 \varphi
-\frac{2}{3 a^2 H^2}{\nabla^2 \varphi}  \right]  ,
\label{measure2}
\end{equation}
and this is the same as Eq.\ (\ref{fosd}), and so the final result is
the same.

Finally, we can check the computation of Ref.\ \cite{Rasanen}, which
used only the first order metric in this gauge. That is, we have to
put $\phi^{(2)}$ and $\psi^{(2)}$ to zero in Eq.\ (\ref{thetatauy}),
\begin{eqnarray}
\theta(t,x^i)&=&
 3H - \frac{2 \nabla^2 \varphi}{3 a^2 H}
-\frac{8 \varphi \nabla^2 \varphi}{9 a^2 H} +
\frac{88 \varphi^{,i} \varphi_{,i} }{45 a^2 H}   -
\frac{4\,{\varphi }^{,j\,k}\,{{\varphi }_{,j\,k}}}
   {9\,\,{a}^4\,{H}^3}+\cdots \nonumber
\\
&=& 3H - \frac{2 \nabla^2 \varphi}{3 a^2 H}
+\frac{128 \varphi^{,i} \varphi_{,i} }{45 a^2 H}
-\frac{8 \partial_i(\varphi \partial^i \varphi)}{9 a^2 H} -
\frac{4\,{\varphi }^{,j\,k}\,{{\varphi }_{,j\,k}}}
   {9\,\,{a}^4\,{H}^3}+\cdots  ,
\end{eqnarray}
and average it with the measure of Eq.\ (\ref{measure2}). Doing so, we
recover the result
\begin{equation}
\langle \theta\rangle (t)= 3H -\frac{22}{45}\frac{\langle
\partial^i\varphi\partial_i\varphi  \rangle}{a^2 H}
+\frac{22}{9} \frac{\langle \partial_i(\varphi \partial^i \varphi) \rangle
}{a^2 H} +\cdots ,
\end{equation}
as in Eq.\ (11) of Ref.\ \cite{Rasanen}.



\end{document}